\documentclass[
  aip,
  jcp,
  reprint,
  twocolumn,
  amsmath,
  amssymb,
  longbibliography
]{revtex4-2}

\usepackage{graphicx}
\usepackage{bm}
\usepackage{booktabs}
\usepackage{xcolor}
\usepackage[colorlinks=true,citecolor=blue,urlcolor=blue,linkcolor=blue]{hyperref}
\hypersetup{
  pdftitle={How long is the coast of a neuron? Resolution-dependent interaction order and computational capacity},
  pdfauthor={Nima Dehghani}
}

\newcommand{\FineConnectedInfo}{0.823}
\newcommand{\ProjectedANDInfo}{0.261}
\newcommand{\ProjectedParityInfo}{0.823}
\newcommand{\FineCategoricalF}{55.0}
\newcommand{\BranchCategoricalF}{96.1}
\newcommand{\SomaCategoricalF}{100.0}

\newcommand{\FineLinearF}{55.0}
\newcommand{\BranchLinearF}{81.3}
\newcommand{\SomaLinearF}{98.9}

\newcommand{\PointMSE}{0.0437}
\newcommand{\ShuffledMSE}{0.0393}
\newcommand{\AlignedMSE}{0.0309}
\newcommand{\EquivalentHidden}{40}
\newcommand{\ScalingInitialMSE}{0.0421}
\newcommand{\ScalingFinalMSE}{0.0152}
\newcommand{\ApproxExponent}{0.31}
\newcommand{\ApproxExponentLow}{0.30}
\newcommand{\ApproxExponentHigh}{0.32}
\newcommand{\RankExponent}{0.24}
\newcommand{\RankExponentLow}{0.23}
\newcommand{\RankExponentHigh}{0.24}

\begin{document}

\title[How Long Is the Coast of a Neuron?]{How Long Is the Coast of a Neuron?\\ Resolution-Dependent Interaction Order and Computational Capacity}

\author{Nima Dehghani}
\affiliation{McGovern Institute for Brain Research, Massachusetts Institute of Technology, Cambridge, Massachusetts, USA}
\affiliation{The NSF AI Institute for Artificial Intelligence and Fundamental Interactions (IAIFI), Cambridge, Massachusetts, USA}
\email{nima.dehghani@mit.edu}
% \date{\today}
% %=============================================
%=============================================
%=============================================
\begin{abstract}
Mandlebrot's coastline paradox reframes measurement as a joint property of an object and its ruler. Here we ask how observational resolution shapes the apparent structure of neural computation. We introduce a scale-indexed decomposition that separates the predictive information captured by direct dependencies from the residual outside that model family. Exact constructions show that projection can convert higher-order dependence into pairwise structure. Projection can increase the direct fraction of surviving information even as total predictive information decreases. A schematic branch-to-soma cascade grounds these effects in local coincidence detection, saturating integration and somatic mixing. Resolving organized nonlinear subunits improves approximation and raises effective representational rank. In a nested hierarchy, these gains occur without increasing information about the target. This distinction separates information availability from access by a restricted readout. A point-neuron or direct population model therefore establishes an effective description at a specified scale. It does not uniquely identify the finer computation that generated the observed relation. Which resolution is adequate depends on the observable and perturbation of interest.
\setcounter{mpfootnote}{1}\footnote{For Context \& overview see:\\ \url{https://neurovium.science/posts/pblog-CoastOfNeuron}}
\end{abstract}


\maketitle
%=============================================
\section{INTRODUCTION}

The question \emph{How long is the coast of Britain?} did more than introduce a memorable fractal~\cite{mandelbrot1967}.  It exposed a dependence that ordinary measurement language tends to hide: the answer belongs jointly to an object and to a ruler.  A coarse ruler crosses bays and omits boundaries that a fine ruler must follow.  The resulting lengths disagree without either measurement being erroneous.

The point neuron is neuroscience's coarse ruler.  From early threshold-unit models onward, a large branch of theoretical neuroscience has represented a cell by a small number of state variables and an input--output rule~\cite{McCullochPitts43,Sompolinsky88,AbbottVanVreeswijk93,vanVreeswijkSompolinsky96,Brunel00}.  This abstraction underlies theories of recurrent dynamics, chaos, balanced excitation and inhibition, and population computation, and has proved extraordinarily productive despite suppressing the spatial structure of the cell.  A related statistical tradition asks how much of population activity can be captured by low-order dependencies alone. In small groups of cells, pairwise maximum-entropy models reproduce surprisingly rich collective activity from first- and second-order statistics~\cite{Schneidman06}. That adequacy is conditional: it is expected in small, sparsely active populations whatever the underlying interactions~\cite{roudi2009}, it gives way to explicit higher-order terms in larger populations under natural stimulation~\cite{ganmor2011}, and it changes with which cells the model retains~\cite{zanoci2019}.

Against this background, a recent maximum-entropy analysis reports striking support for the coarse description at the level of the single cell: direct, equal-time dependencies on relatively few recorded neurons explain more than $90\%$ of output entropy in mouse recordings and 62\% in \textit{C.\ elegans}~\cite{lynn2026}.  The analysis is careful about what is measured.  Its inferred weights are functional dependencies among recorded cells, not necessarily synapses, and unexplained variability can combine higher-order dependence, temporal dependence, unobserved inputs, and stochasticity.

We propose a scale-theoretic fromalism.  The success of a direct model at the resolution of a population recording characterizes an effective relation among the variables that survive measurement and projection.  It does not imply that the cellular computation realizing that relation is itself direct, additive, or single-stage.  This distinction matters because a neuron is not, at finer resolution, a single computational degree of freedom.  Dendrites contain spatially distributed nonlinear subunits~\cite{london2005,polsky2004,gidon2020}, and the input--output transformation of a detailed neuron can require a multilayer artificial network for accurate functional emulation~\cite{poirazi2003,beniaguev2021}. Thus, the population-level and dendritic descriptions are not competing accounts of the same object; they are descriptions at different observational scales, connected by a many-to-one projection.  The question is therefore not whether the reduced model is useful, but whether predictive adequacy at one scale warrants identifying its effective dependencies with the computation that produces them at another.

The coastline analogy is deliberately epistemic.  We do not claim that every dendrite possesses an exact fractal dimension over an unlimited range.  Neuronal arbors do exhibit scale-dependent geometric organization and constrained branching~\cite{smith2021,cuntz2010,cuntz2012}, but geometry is not the premise of the argument.  What matters is that observation can collapse many underlying degrees of freedom into the same measured variable while nonlinear computation persists below that resolution. % Nor is the aim to replace the reduced neuron with a realistic reconstruction or a full compartmental simulation.  Doing so would answer a different question.

We instead ask how much of a neuron's computation is resolved at a given observational scale, which distinctions are erased by projection, and which reappear as the ruler is refined.  We approach this problem through an information-theoretic lens together with constructive counterexamples in which the efficiency of neural-network architectures depends on the representation available at each scale.  These examples show that the same coarse input--output relation can be compatible with markedly different internal computations, even when the reduced description is highly predictive.  The coastline problem for a neuron is therefore not whether computation is visible or hidden, but how its apparent structure changes with the scale at which it is observed.

\section{METHODS $\&$ RESULTS}
\subsection{What a Direct Model Measures}

Let $\bm X_\varepsilon=(X_{\varepsilon,1},\ldots,X_{\varepsilon,m})$ denote the effective inputs visible with ruler $\varepsilon$, and let $Y\!\in\!\{0,1\}$ be the output.  Larger $\varepsilon$ denotes a coarser description.  At each resolution, define the direct conditional family
\begin{equation}
Q_{\bm\theta,\varepsilon}(y=1\mid\bm x)
=\sigma\!\left[b+\sum_i u_i(x_i)\right],
\label{eq:direct-family}
\end{equation}
where each $u_i$ is an arbitrary one-coordinate function.  For binary inputs this reduces to the familiar logistic model $b+\sum_iw_ix_i$.  The generalized form becomes important when coarse variables possess several distinct states: assigning three categorical states the numerical labels $0,1,2$, for example, should not impose an artificial linear ordering on their effects.  Allowing an independent value of $u_i$ for each state makes direct dependence invariant to such arbitrary categorical coding.

Let $Q_\varepsilon^\star$ be the information projection of the true conditional distribution onto Eq.~(\ref{eq:direct-family}).  Using base-two logarithms, define
\begin{align}
S_{\rm tot}(\varepsilon)&=H(Y),\ \\
S_{\rm true}(\varepsilon)&=H(Y\mid\bm X_\varepsilon),\ \\
S_{\rm dir}(\varepsilon)&=-\mathbb E_{P_\varepsilon}
\log_2 Q_\varepsilon^\star(Y\mid\bm X_\varepsilon).
\label{eq:entropies}
\end{align}
Here $S_{\rm tot}$ is the uncertainty of the output before observing the inputs, $S_{\rm true}$ is the irreducible conditional uncertainty at resolution $\varepsilon$ (explicitly allowing for all synergistic and higher-order interactions present in the true distribution $P_{\epsilon}$), and $S_{\rm dir}$ is the uncertainty remaining under the best additive-logit description at that resolution.  The last quantity is a conditional cross-entropy.  Moment matching connects it to the maximum-entropy expression (like what was used in Ref.~\cite{lynn2026}), while the cross-entropy form makes the decomposition explicit:
\begin{align}
I_{\rm avail}
&= S_{\rm tot}-S_{\rm true}
= I(Y;\bm X_\varepsilon),
\label{eq:IavailDiff}\ \\
I_{\rm dir}
&= S_{\rm tot}-S_{\rm dir},
\label{eq:Idir}\ \\
I_{\rm higher}
&= S_{\rm dir}-S_{\rm true}
\label{eq:Ihigher}\ \\
&= \mathbb E_{\bm X_\varepsilon}
D_{\rm KL}\left[
P_\varepsilon(Y\mid\bm X_\varepsilon)
\,\Vert\,
Q_\varepsilon^\star(Y\mid\bm X_\varepsilon)
\right]. 
\label{eq:Ihigher_KL}
\end{align}
This decomposition relies on the algebraic difference between the direct model's cross-entropy and the true conditional entropy in Eq.~\eqref{eq:Ihigher}. Expanding these logarithmic terms reveals that this difference is mathematically identical to the conditional Kullback--Leibler (KL) divergence. By combining the expectations, the subtraction of $-\mathbb{E}_{P_{\varepsilon}}\log_{2}P_{\varepsilon}$ from $-\mathbb{E}_{P_{\varepsilon}}\log_{2}Q_{\varepsilon}^{\star}$ yields $\mathbb{E}_{P_{\varepsilon}}\log_{2}(P_{\varepsilon}/Q_{\varepsilon}^{\star})$, which is precisely the conditional divergence in Eq.~\eqref{eq:Ihigher_KL}.

Thus $I_{\rm higher}$ is precisely the conditional information missed by the optimal direct family.  It measures non-additive conditional structure relative to that family rather than a particular notion of information-theoretic synergy.

By Gibbs' inequality, KL divergence is non-negative, so
$S_{\rm dir}\geq S_{\rm true}$.
The direct family also contains the intercept-only model $Q(Y)=P(Y)$, so optimality of $Q_\varepsilon^\star$ gives
$S_{\rm dir}\leq S_{\rm tot}$.  Hence
\begin{align}
I_{\rm avail} &= I_{\rm dir} + I_{\rm higher}, 
\qquad
S_{\rm tot} \geq S_{\rm dir} \geq S_{\rm true}.
\label{eq:S_ineq}
\end{align}
The available information therefore partitions exactly into the portion captured by the direct model and a non-negative interaction residual.

Two resultant ratios answer different questions:
\begin{align}
R(\varepsilon)&=\frac{I_{\rm dir}}{S_{\rm tot}},
\label{eq:FracR}\ \\
F(\varepsilon)&=\frac{I_{\rm dir}}{I_{\rm avail}}
=1-\frac{I_{\rm higher}}{I_{\rm avail}},
\qquad(I_{\rm avail}>0).
\label{eq:FracF}
\end{align}
$R$ measures the predictive information captured by the direct model relative to the total uncertainty of the output.  $F$, by contrast, asks what fraction of the information available at the chosen resolution is captured by the direct family.  Since $I_{\rm avail}\leq S_{\rm tot}$, we have $R\leq F$ whenever $I_{\rm avail}>0$. 

This distinction between $R(\varepsilon)$ and $F(\varepsilon)$ is central under coarse-graining. Projection cannot increase $I_{\rm avail}$, but it can alter how the surviving information is distributed between $I_{\rm dir}$ and $I_{\rm higher}$. If non-additive structure is discarded more rapidly than direct structure, $F$ can approach unity even while the total predictive information decreases. Consequently, $F$ can rise toward 100\% even as the total-entropy ratio $R$ falls. A high direct-information fraction $F$ at a coarse scale therefore does not prove that the underlying biological mechanism is simple; it indicates only that little of the information surviving at that ruler lies outside the direct family. The coarse ruler may be blind to the higher-order interactions that are erased by projection.
 
 The reported quantity in Ref.~\cite{lynn2026} is our $R$, the fraction of total output entropy explained by direct dependencies. The distinction between $R$ and $F$ will be important below.

\subsection{What Coarse-Graining Guarantees}

Suppose that a coarser observable is obtained from the variables at resolution $\varepsilon$ through
$\bm Z=T(\bm X_\varepsilon)$.  The data-processing inequality gives
\begin{equation}
I(Y;\bm Z)\leq I(Y;\bm X_\varepsilon).
\label{eq:dpi}
\end{equation}
Thus coarse-graining cannot increase the total predictive information available about $Y$.  In the notation of the previous section, $I_{\rm avail}$ can only decrease or remain unchanged as information is discarded by the projection.

It is important, however, to separate this absolute information loss from changes in the structure of the surviving dependence.  Equation~\eqref{eq:dpi} does not, by itself, order either the higher-order residual $I_{\rm higher}$ or the normalized direct fraction $F$.  The reason is that the optimal direct model is defined relative to the observable coordinates themselves, and that admissible family changes when $\bm X_\varepsilon$ is replaced by $\bm Z$.  A blanket claim that higher-order information must decrease under every coarse-graining would therefore be false.

A stronger statement becomes possible under an additional closure condition.  Form the joint direct surrogate
\begin{equation}
Q^\star_\varepsilon(y,\bm x)
=
P_\varepsilon(\bm x)
Q^\star_\varepsilon(y\mid\bm x),
\label{eq:jointsurrogate}
\end{equation}
which preserves the true distribution of the observed inputs while replacing the conditional response by its optimal direct approximation.  Project both the true joint distribution $P_\varepsilon$ and the surrogate in Eq.~\eqref{eq:jointsurrogate} through the same map $T$.  If the conditional distribution induced by the projected surrogate remains within the direct family defined at the coarse scale, then the family is closed under that projection.

Under this condition, two inequalities apply in sequence.  First, because the direct model is refitted after coarse-graining, the optimal coarse direct model can perform at least as well as the projected fine-scale surrogate.  Second, KL divergence contracts when both distributions are passed through the same stochastic map.  Together these facts give

\begin{align}
 I_{\rm higher}(\bm Z)
 &\leq D_{\rm KL}(P_{Y,\bm Z}\Vert Q^\star_{Y,\bm Z})\nonumber\\
 &\leq D_{\rm KL}(P_{Y,\bm X}\Vert Q^\star_{Y,\bm X})
 =I_{\rm higher}(\bm X).
 \label{eq:closure}
\end{align}

The first inequality follows from optimization within the coarse direct family; the second is KL contraction.  Equation~\eqref{eq:closure} therefore provides a sufficient condition under which the absolute non-additive residual cannot increase with coarsening.  The closure condition is essential: without it, projection can change the form of the admissible direct representation, and no universal ordering of $I_{\rm higher}$ follows.

Even when Eq.~\eqref{eq:closure} holds, however, the normalized direct fraction need not increase.  From Eq.~\eqref{eq:FracF}, we have
\begin{equation}
F
=
1-\frac{I_{\rm higher}}{I_{\rm avail}},
\end{equation}
so coarse-graining acts on both the residual in the numerator and the total available information in the denominator.  $F$ moves toward unity only when $I_{\rm higher}$ is lost proportionally faster than $I_{\rm avail}$. If the two are lost at the same rate, or the available information is lost faster, $F$ is unchanged or falls.  Because the denominator moves with the numerator, $F$ carries no information about how much predictive information survives.  We call the \emph{denominator trap} the error of reading a high $F$ as evidence of a simple mechanism without checking $I_{\rm avail}$: $F$ can approach unity precisely because almost everything, interactions included, has been discarded.

The data-processing inequality therefore guarantees loss, or at most preservation, of total predictive information; closure of the direct family additionally guarantees contraction of the absolute higher-order residual, $I_{\rm higher}(\bm Z)\leq I_{\rm higher}(\bm X)$. Neither result mandates that a coarse description will appear increasingly direct.  That final behavior depends on how the particular projection redistributes the information that survives between $I_{\rm dir}$ and $I_{\rm higher}$.  Whether coarse-graining simplifies the apparent computation is consequently a property of the measurement map and the underlying dependency structure, and must be evaluated scale by scale.

\subsection{Interaction Order Changes Under Projection}

Coarse-graining can change not only how much information is retained, but also the order at which the surviving dependence appears.  A dependency that is irreducibly higher-order in one set of observables can become an ordinary pairwise dependency after projection.  This distinction is important for mechanistic inference: observing pairwise structure at a coarse scale does not establish that the process generating it is pairwise at finer scales.

A three-spin construction makes this ambiguity exact.  Let $s_i\in\{-1,+1\}$, then the joint probability distribution is
\begin{equation}
P_K(\bm s)=\frac{\exp(Ks_1s_2s_3)}{8\cosh K}.
\label{eq:three-spin}
\end{equation}
The distribution in Eq.~\eqref{eq:three-spin} contains only a three-body coupling, with $K$ setting the coupling strength. Because the interaction depends on the triad product $s_1s_2s_3$, marginalizing out any single spin cancels the coupling term and leaves the remaining pair completely uncorrelated. Therefore, every one- and two-spin marginal is uniform, and the pairwise maximum-entropy distribution $\widetilde P^{(2)}$ is uniform over the eight possible states. Its entropy is exactly $3$ bits. The connected information of order three is therefore exactly the difference between the entropy of three independent spins and the true entropy of the coupled system\cite{schneidman2003,amari2001}
\begin{equation}
I^{(3)}=H[\widetilde P^{(2)}]-H[P_K]=3-H[P_K].
\label{eq:connected-three}
\end{equation}
At $K=1.8$, exact enumeration gives $I^{(3)}=\FineConnectedInfo$ bits.  Thus the departure from the pairwise maximum-entropy description is entirely third-order in the fine variables.

The interaction order changes when the same distribution is viewed through a different ruler. To demonstrate how this interaction order shifts under coarse-graining, we write $b_i=(s_i+1)/2$ and retain only $x=b_1$ and a compressed variable $z=b_2\wedge b_3$. Because $x$ and $z$ depend on disjoint fine coordinates, they are independent at $K=0$ (zero coupling, where Eq.~\eqref{eq:three-spin} is uniform). This provides a clean baseline: any mutual information observed at nonzero coupling reflects the coarse-grained effect of the triadic interaction, rather than a spurious dependence introduced by overlapping projections.

\begin{figure*}
\centering
\includegraphics[width=0.92\linewidth]{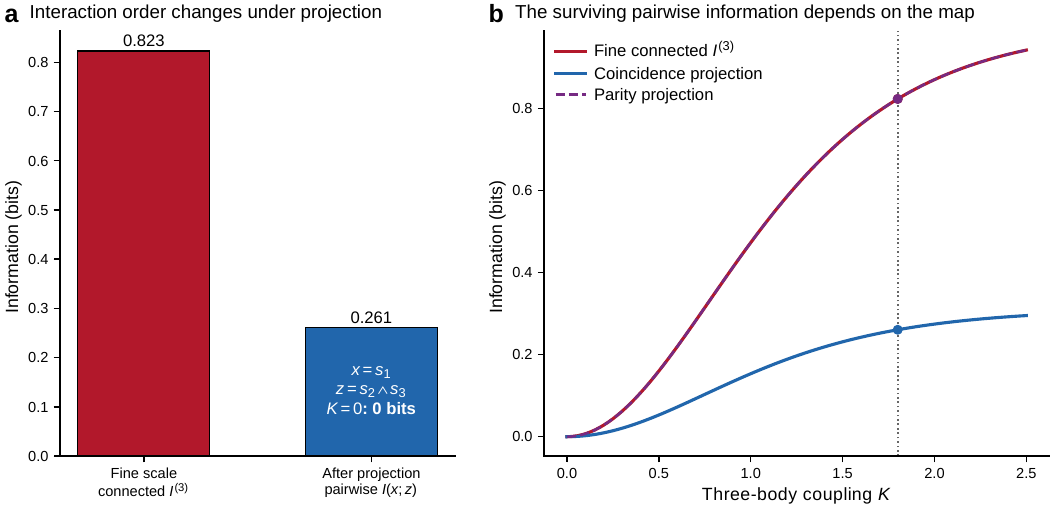}
\caption{\label{fig:projection}\textbf{A higher-order dependency becomes pairwise after projection.}
(a) Exact enumeration at $K=1.8$ gives \FineConnectedInfo\ bits of connected third-order information at the fine scale and \ProjectedANDInfo\ bits of pairwise mutual information after the disjoint coincidence projection $x=b_1$, $z=b_2\wedge b_3$.  The projected variables are independent at $K=0$.  (b) The information that survives as pairwise depends on the map: an XNOR/parity projection retains all of $I^{(3)}$, whereas the coincidence map retains only part.}
\end{figure*}

At $K=1.8$, however, the ordinary mutual information between these same projected variables is \ProjectedANDInfo\ bits (Fig.~\ref{fig:projection}).  Part of the third-order dependence quantified by Eq.~\eqref{eq:connected-three} has therefore become an ordinary pairwise dependency after projection.  No new information has been created; the many-to-one map has changed the order at which part of the existing statistical structure is represented.

The dependence on the ruler is even clearer under a second projection.  If instead $z=\mathbf{1}[b_2=b_3]$, the projected pairwise information is \ProjectedParityInfo\ bits, equal to the full value of $I^{(3)}$ in Eq.~\eqref{eq:connected-three}.  This occurs because the equality variable encodes the product $s_2s_3$. The fine-scale interaction $s_1s_2s_3$ is thereby represented at the coarse scale as a dependency between two observables.  Under this projection, the entire third-order dependence survives, but its apparent interaction order has changed from three to two.

The same fine distribution can thus exhibit no pairwise dependence in its original coordinates, a weaker pairwise dependence under one projection, and the full dependence as a pairwise relation under another.  Interaction order is therefore not invariant under a many-to-one measurement map.  It is a property of the statistical structure relative to the observables through which that structure is resolved.  A strong pairwise dependency at coarse resolution cannot, by itself, identify a pairwise mechanism at finer resolution. The ruler can change the apparent order of the computation.

\subsection{The Branch-To-Soma Cascade}

Having established that interaction order can change under mathematical projection, we now translate the same logic into a neuron-like architecture. The purpose is not to construct a compartmental model of a particular dendrite. It is to show, with an exactly enumerable example, how familiar biophysical operations---local coincidence detection, saturating integration, and somatic mixing---can act as successive coarse-graining maps. We therefore follow the same signal through three observational rulers. Individual synaptic contacts, two dendritic branches, and a single somatic variable. The underlying input--output law is fixed throughout. What changes is the set of distinctions available to the observer, and under coarse‑graining a neuron with genuinely synergistic substructure can masquerade as an additive unit.

\subsubsection{The fine-scale ground truth: NMDA coincidence detection}

We begin at the finest resolution of the construction. Six independent binary inputs, $\bm x\in\{0,1\}^{6}$, are divided evenly between two three-input subunits. Think of a neuron receiving six independent binary inputs, distributed evenly across two dendritic branches. In this construct, the output probability is governed strictly by third-order  interactions of the two input branches:
\begin{equation}
P(Y=1\mid\bm x)=0.06+0.40
(x_{1}x_{2}x_{3} + x_{4}x_{5}x_{6}).
\label{eq:cascade}
\end{equation}
Thus the baseline firing probability is $0.06$, one active triplet raises it to $0.46$, and simultaneous activation of both triplets gives the maximum probability $0.86$. All $2^6=64$ input configurations can therefore be enumerated exactly.

% NMDA-mediated dendritic plateau potentials. Because the NMDA receptor channel is blocked by a magnesium ion at resting potentials, a single synaptic input cannot dislodge the block. It requires the tight spatiotemporal coincidence of multiple clustered inputs to eject the magnesium and trigger a nonlinear calcium influx.
The two products in Eq.~\eqref{eq:cascade} are deliberately sharp abstractions of local dendritic coincidence detection. NMDA-dependent dendritic events provide the biological motivation: because NMDA conductance is strongly voltage dependent through extracellular Mg$^{2+}$ block, sufficiently coordinated local excitation can recruit a regenerative response that isolated inputs do not produce \cite{london2005,polsky2004,palmer2014}. The number three is not intended as a literal universal threshold for an NMDA event. Its role is mathematical: $x_{1}x_{2}x_{3}$ is nonzero only when all three members of the local group are active, providing the simplest exact representation of a third-order coincidence.

\begin{figure*}
\centering
\includegraphics[width=0.95\linewidth]{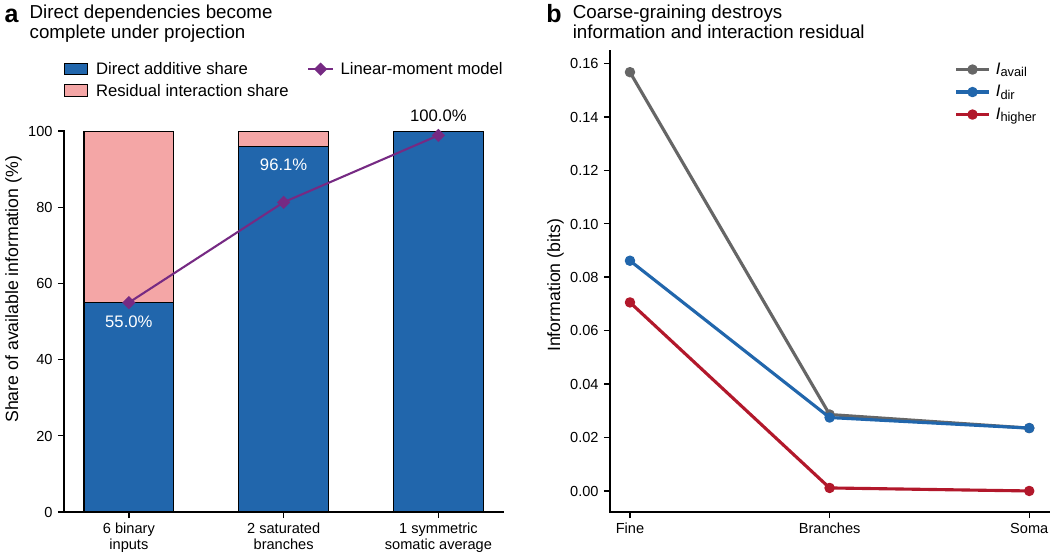}
\caption{\label{fig:cascade}\textbf{Direct dependencies become more complete along an exact cascade.}
(a) Blue bars show $F$ for the coordinate-invariant categorical-additive family; pink shows the residual share.  Purple diamonds show the more restrictive linear-moment family, which yields \FineLinearF\%, \BranchLinearF\% and \SomaLinearF\%.  (b) Coarse-graining reduces the available information $I_{\rm avail}$, the captured direct information $I_{\rm dir}$ and, more sharply, the conditional-KL residual $I_{\rm higher}$.  Values are exact sums over all 64 input states.}
\end{figure*}

At this contact-level ruler, the generative law is therefore explicitly organized around third-order products. This does not mean that individual contacts carry no predictive information. Marginalizing over the other inputs produces direct one-coordinate effects that an additive model can capture. What it cannot reconstruct is the full dependence on the joint configuration. For the coordinate-invariant categorical-additive family defined in Eq.~\eqref{eq:direct-family}, the direct fraction is consequently only $55.0\%$. In other words, slightly more than half of the available information can be assigned to direct single-coordinate effects, while the remainder requires their joint configuration.

For continuity with the scalar logistic convention, the more restrictive linear-moment model gives the sequence
\begin{equation}
 F:\quad
 \FineLinearF\%\ \longrightarrow\
 \BranchLinearF\%\ \longrightarrow\
 \SomaLinearF\%.
\label{eq:cascade-linear}
\end{equation}
The first entry, $55.0\%$, is therefore the fine-scale reference against which the subsequent projections are measured.

\subsubsection{Intermediate projection: sublinear summation and saturation}

We next change the ruler to an intermediate macroscopic one. Suppose that individual contacts are no longer resolved and that the observer instead has access to one effective variable from each dendritic branch. We define those branch observables by the saturated sums
\begin{equation}
B_1=\min(x_1+x_2+x_3,2),\quad
B_2=\min(x_4+x_5+x_6,2),
\label{eq:branch-observables}
\end{equation}
so that each branch can now occupy only the states $0$, $1$, or $2$.

The hard ceiling in Eq.~\eqref{eq:branch-observables} is a stylized representation of saturating or sublinear local integration, not a quantitative membrane equation. Real dendritic responses can become sublinear through several conductance and driving-force mechanisms\cite{london2005}. As a local dendritic membrane depolarizes toward the reversal potential of its excitatory ion channels, the electrochemical driving force decreases, and voltage summation plateaus. However, active dendrites can also exhibit strongly supralinear regimes \cite{polsky2004,gidon2020}. What matters for the present construction is the many-to-one character of the map. Explicitly, distinct microscopic input configurations become indistinguishable once only $B_1$ and $B_2$ are retained.

The state collapse can be seen directly. If exactly two contacts on the first branch are active, then $B_1=\min(2,2)=2$. If all three are active, $B_1=\min(3,2)=2$ as well. The fine-scale states $(1,1,0)$ and $(1,1,1)$ are therefore assigned the same observed branch value even though Eq.~\eqref{eq:cascade} treats them differently. The triplet contribution is zero in the first case and nonzero in the second. The projection has removed precisely the distinction needed to determine whether the local third-order gate was completed.

That lost distinction reduces absolute predictive information. The available information falls from $I_{\rm avail}=0.156742$ bits at the six-contact ruler to $0.028607$ bits at the two-branch ruler. But the higher-order residual falls much more sharply, from $I_{\rm higher}=0.0706$ to $0.0011$ bits. Consequently, the information that remains after projection is overwhelmingly describable by direct effects of the two observed branch variables. For the categorical-additive family, $F$ rises from $55.0\%$ to $96.1\%$.

This is the important direction of the effect. The branch projection has not made the microscopic triplet interaction disappear from the generative mechanism. It has erased much of the observer's ability to distinguish the microscopic states on which that interaction depends. The coarse conditional distribution therefore looks substantially more direct than the fine one.

\subsubsection{Somatic projection: aggregation and the denominator trap}

The final ruler retains only a single somatic observable,
$Z=(B_1+B_2)/2$. The nine possible branch pairs $(B_1,B_2)$ are thereby compressed into only five somatic values,
$Z\in\{0,0.5,1.0,1.5,2.0\}$. Relative to the original $64$ synaptic configurations, this is an extreme contraction of the observable state space.

The symmetric average performs a second kind of coarse-graining in addition to the within-branch saturation of Eq.~\eqref{eq:branch-observables}. It removes branch identity and the spatial origin of the input. For example, $(B_1,B_2)=(2,0)$ gives $Z=(2+0)/2=1$, while $(B_1,B_2)=(0,2)$ gives exactly the same somatic value. The observer can therefore no longer determine which branch generated a given contribution to the soma.

The absolute information measures make clear how much has been discarded. Across the full cascade,
$I_{\rm avail}$ falls from $0.156742$ to $0.028607$ to $0.023526$ bits. The categorical higher-order residual simultaneously falls from $0.0706$ to $0.0011$ to $0.0000$ bits (Fig.~\ref{fig:cascade}). The direct-information fraction for the categorical-additive family is therefore
\begin{equation}
 \FineCategoricalF\%\ \longrightarrow\
 \BranchCategoricalF\%\ \longrightarrow\
 \SomaCategoricalF\%.
\label{eq:cascade-categorical}
\end{equation}
That is, the same underlying system moves from $55.0\%$ direct at contact resolution, to $96.1\%$ at branch resolution, to $100.0\%$ at somatic resolution.

The last number requires careful interpretation. At the somatic ruler there is only one observed coordinate, $Z$. The categorical direct family of Eq.~\eqref{eq:direct-family} allows an arbitrary one-coordinate function $u(Z)$, so there can be no residual interaction \emph{among observed coordinates} (i.e. there is only one coordinate left). Thus $I_{\rm higher}=0$ and $I_{\rm dir}=I_{\rm avail}$ exactly. The $100\%$ result does not mean that the six-contact mechanism in Eq.~\eqref{eq:cascade} has become microscopically additive. It means that, after the projection has removed the relevant internal distinctions, all of the information still accessible at the somatic ruler is representable as a direct dependence on the single retained variable.

This is the denominator trap in its sharpest form. The fraction $F=I_{\rm dir}/I_{\rm avail}$ reaches unity while the denominator itself contracts by $85\%$. The complementary total-entropy ratio makes that contraction visible: $R$ is only $13.6\%$ at the contact scale, $4.3\%$ at the branch scale, and $3.7\%$ at the soma. A coarse model can therefore be perfect relative to the information remaining at its own resolution while retaining only a small fraction of the predictive structure that existed before projection. High $F$ and low $R$ are not contradictory statements; they answer different questions. These numerical ratios are not intended as a model of the recordings in Ref.~\cite{lynn2026}; the cascade is a constructive example showing how projection can increase apparent directness while the predictive information retained by the coarse representation decreases.

The use of the categorical-additive family ($\sum w_{i} \cdot \mathbf{1}[B=i]$) is also important here. If the numerical labels $0$, $1$, and $2$ assigned to a branch state were forced to enter a straight-line predictor, an apparent residual could arise simply because the chosen coordinate happened to be nonlinearly encoded. Allowing an arbitrary one-coordinate function, as in Eq.~\eqref{eq:direct-family}, removes that coordinate artifact. The collapse from $55.0\%$ to $96.1\%$ to $100.0\%$ therefore reflects the projection itself rather than failure of an unnecessarily restrictive linear parameterization.

\subsubsection{Control: asymmetric mixing via electrotonic attenuation}

The symmetric average is intentionally a strong many-to-one projection, so the result should not be mistaken for a claim that every somatic mixing rule produces the same conversion. To test the dependence on the projection itself, we replace the symmetric mixture by
$Z=0.6B_1+0.4B_2$. Unequal weighting is biologically natural as an effective description because dendritic signals undergo location- and state-dependent electrotonic transfer before reaching the soma \cite{london2005}. A branch physically closer to the soma will experience less electrotonic attenuation than a distal branch. The particular $0.6/0.4$ values are not intended as a fit to a specific morphology; they provide a controlled perturbation of the projection geometry.

The asymmetry removes the degeneracy generated by symmetric averaging. The branch states $(2,0)$ and $(0,2)$, which both produced $Z=1$ under equal mixing, now produce
$0.6(2)+0.4(0)=1.2$ and
$0.6(0)+0.4(2)=0.8$, respectively. The nine branch states now map to nine distinct somatic values, so branch identity is recoverable from the scalar somatic coordinate.

For this control, the relevant comparison is the linear-moment family. The fully categorical family is uninformative for distinguishing two one-dimensional scalar projections, because an arbitrary function of a single scalar variable is direct by construction. The linear family instead asks whether the remaining input--output dependence is simple in the particular numerical coordinate produced by the projection. At the fine quantization used in Fig.~\ref{fig:projection-resolution}, changing the somatic mixture from $0.5/0.5$ to $0.6/0.4$ lowers the linear-family direct fraction $F$ from $98.9\%$ to $78.4\%$.

The control therefore exposes the role of the map itself. The microscopic law in Eq.~\eqref{eq:cascade} has not changed, nor have the six input variables. Only the projection from branch states to the measured somatic coordinate has changed. Yet the apparent directness of the resulting description changes substantially. Directness is consequently not an intrinsic label that can be attached to the neuron independently of resolution. It belongs jointly to the underlying distribution, the coordinates retained by the observer, the model family used to describe those coordinates, and the biophysical projection that connects one ruler to the next.

\subsection{From One Weight To A Field Of Contacts}

The branch-to-soma cascade showed how spatial aggregation can make a neuron appear increasingly direct even when its fine-scale input--output law contains explicit higher-order interactions. We now apply the same argument to a more elementary object in network models: the synaptic weight. The question is what a single edge $w_{ij}$ represents when the biological connection that it summarizes is resolved at progressively finer spatial scales.

At point-neuron resolution, one connection is naturally represented by one effective coefficient. At contact resolution, however, the same anatomical relation can contain multiple spatially embedded synaptic events whose influence depends on where they arrive and on the local state into which they arrive. The purpose of this section is therefore not to deny the usefulness of a scalar weight. It is to identify what has already been integrated out when a spatially distributed, state-dependent biological connection is represented by one number.

\subsubsection{The scalar-weight abstraction}

In a standard point-neuron description, the influence from presynaptic cell $j$ to postsynaptic cell $i$ is represented by a scalar coefficient $w_{ij}$. Depending on the model, that coefficient may multiply an instantaneous activity variable, a filtered spike train, or another effective presynaptic signal. In each case, however, the spatial structure of the connection inside neuron $i$ has been removed from the representation. The model retains one edge between two neuronal nodes rather than the individual sites at which that edge is physically realized.

At a finer microscopic ruler $\varepsilon$, the appropriate object need not be a single scalar. It can instead be represented as a collection of spatially embedded contacts,
\begin{equation}
\mathcal W_{ij,\varepsilon}
=
\left\{
w^{(k)}_{ij}(\bm r;\varepsilon)
\right\}_{k=1}^{n_{ij}(\varepsilon)}.
\label{eq:contact-field}
\end{equation}
where $n_{ij}(\varepsilon)$ denotes the number of contacts resolved between cells $j$ and $i$, $k$ indexes an individual contact, and $\bm r$ specifies its location within the postsynaptic arbor. Equation~\eqref{eq:contact-field} is not intended as a detailed synaptic model. It expresses the change in ontology produced by changing the ruler. Through this lens, what was one edge at neuronal resolution becomes a set of spatially distinguished degrees of freedom at contact resolution.

The distinction matters because dendritic location is computationally consequential. Synaptic efficacy at the soma depends on electrotonic distance, local membrane conductances, neighboring excitation and inhibition, and the nonlinear state of the dendritic compartment through which the signal propagates~\cite{london2005,polsky2004,poirazi2003}. Thus two contacts associated with the same presynaptic neuron need not participate in the same local computation merely because a coarse graph assigns them to the same edge.

\subsubsection{Context-dependent influence and higher-order structure}
Consider one resolved contact $k$ from neuron $j$ onto a dendritic branch of neuron $i$. Its contribution cannot in general be specified from presynaptic activity alone. If the local branch is relatively hyperpolarized and away from a conductance or voltage-saturation regime, activation of the contact can produce one incremental response. If coincident activity elsewhere has already altered the local voltage, recruited active conductances, or changed the balance of excitation and inhibition, activation of the same contact can produce a substantially different response \cite{london2005,polsky2004,gidon2020} (e.g. the same spike from cell $j$ at synapse $k$ may generate zero additional depolarization).

The relevant point is therefore not simply that synaptic strength varies. It is that the effect assigned to one contact can depend on the state created by other contacts. Let $\mathcal N_\varepsilon$ denote this fine-scale context: local membrane voltage, active conductance state, nearby synaptic activity, and other unresolved variables relevant at ruler $\varepsilon$. The contribution associated with $w_{ij}^{(k)}$ is then conditional on $\mathcal N_\varepsilon$ rather than being an independent term whose effect is fixed across states.

This dependence can be stated directly in the language of the direct-family model introduced in Eq.~\eqref{eq:direct-family}. Let $x_k$ denote the state of contact $k$ and let $n$ denote the resolved local context. Define the context-dependent log-odds increment
\begin{equation}
\begin{aligned}
\Delta_k(n)=
\operatorname{logit}P(Y=1\mid x_k=1,n) \\
-\operatorname{logit}P(Y=1\mid x_k=0,n).
\label{eq:contact-context}
\end{aligned}
\end{equation}
For the additive direct family in Eq.~\eqref{eq:direct-family}, $\Delta_k(n)$ is independent of $n$: the contribution assigned to $x_k$ cannot change with the state of another coordinate. If instead $\Delta_k(n)\neq\Delta_k(n')$ for two accessible local states $n$ and $n'$, then the conditional response contains an interaction that cannot be represented by an independent one-coordinate contribution from $x_k$ alone.

Dependence of the conditional law on $n$ does not by itself imply $I_{\rm higher}>0$: such dependence may still be represented by a separable term $u_n(n)$. The stronger condition in Eq.~\eqref{eq:contact-context} is that the increment attributable to $x_k$ itself changes with context. If $\Delta_k(n)\neq\Delta_k(n')$, no representation of the form $u_k(x_k)+u_n(n)$ can reproduce both increments. The true conditional law then lies outside the direct family, and by the conditional-KL representation in Eq.~\eqref{eq:Ihigher_KL}, the optimal direct approximation retains a positive residual, $I_{\rm higher}>0$.

This distinction gives a precise information-theoretic meaning to local dendritic context. A contact whose effective contribution is invariant across the surrounding state can remain informationally direct even if its absolute magnitude differs across cells or locations. A contact whose contribution changes according to the configuration of other resolved variables instead participates in a joint dependence. The latter is the fine-scale structure that disappears when the local context is removed from the observer's coordinate system.

This is the microscopic version of the branch example in Section V. At the coarse ruler, neuron $j$ may appear to exert one direct influence on neuron $i$. At the fine ruler, that influence can be distributed across contacts whose effects depend jointly on their dendritic location and on the activity surrounding them. The anatomical connection remains monosynaptic; the informational decomposition need not remain additive.

\subsubsection{The point neuron as a coarse-graining map}

A point-neuron model does not observe $\bm r$, the individual elements of $\mathcal W_{ij,\varepsilon}$, or the local state $\mathcal N_\varepsilon$. It observes variables defined at neuronal resolution and seeks an effective dependency between them. The corresponding somatic weight can therefore be viewed as the result of a coarse-graining projection,
\begin{equation}
w_{ij}^{\rm soma}
=\Pi_\varepsilon[\mathcal W_{ij,\varepsilon},\mathcal N_\varepsilon].
\label{eq:soma-weight}
\end{equation}
Equation~\eqref{eq:soma-weight} is deliberately schematic. The operator $\Pi_\varepsilon$ stands for everything involved in passing from the resolved contact description to the effective somatic one: dendritic integration, electrotonic propagation, marginalization over unresolved local states, measurement, and finally the effective parameterization used by the coarse model.

This makes clear what the scalar weight does and does not mean. The quantity $w_{ij}^{\rm soma}$ need not correspond to any individual member of the contact field in Eq.~\eqref{eq:contact-field}. It summarizes their net predictive influence after the fine variables have been integrated out. Different microscopic configurations of contacts and local states can therefore produce the same effective scalar, just as different synaptic configurations in the branch-to-soma cascade produced the same coarse observable.

To connect this projection to the information hierarchy, collect the
resolved variables into
$X_\varepsilon=(\mathcal W_{ij,\varepsilon},\mathcal N_\varepsilon)$
and define the coarse variable
$X_{\rm soma}=\Pi_\varepsilon(X_\varepsilon)$.  Because
$X_{\rm soma}$ is a deterministic function of the fine state, the
data-processing inequality gives
\begin{equation}
 I(Y;X_{\rm soma})\leq I(Y;X_\varepsilon).
 \label{eq:contact-dpi}
\end{equation}
Thus the available information itself can only contract under this
projection.  Under the closure condition of Eq.~\eqref{eq:closure},
the absolute higher-order residual contracts as well,
$I_{\rm higher}(\bm Z)\leq I_{\rm higher}(\bm X_\varepsilon)$. The scalar weight in Eq.~\eqref{eq:soma-weight} can therefore be read
as an explicit realization of the coarse-graining map analyzed in
Section III.

From the perspective of the information hierarchy, this projection can remove precisely the distinctions required to expose context-dependent interactions. Structure that was visible as a dependence among resolved contacts may become absorbed into an effective direct dependency between neuronal variables, while variation associated with unresolved fine states appears as conditional variability at the coarse scale. The scalar weight is consequently an effective parameter of the chosen ruler, not a literal inventory of the microscopic operations implementing the connection.

This is also why the success of the coarse description should not be interpreted as evidence against the finer one. A fitted point-neuron model may predict somatic activity extremely well precisely because the variables that distinguish alternative fine-scale implementations have been marginalized over. Predictive adequacy of $w_{ij}^{\rm soma}$ establishes the usefulness of the projected edge for the declared observable. It does not establish that $\mathcal W_{ij,\varepsilon}$ contained only one computational degree of freedom before projection.

\subsubsection{Anatomical vs.\ informational directness}

The distinction exposes a conceptual trap that recurs throughout computational neuroscience and connectionist network models: anatomical directness and informational directness answer different questions. Anatomical directness asks whether cell $j$ forms a direct connection onto cell $i$. Informational directness asks whether the dependence of the output on the variables resolved by the observer can be represented without irreducible joint terms.

The distinction can be stated formally. Let $A_{ij}=1$ denote the existence of an anatomical connection from neuron $j$ to neuron $i$, so that at contact resolution $A_{ij}=1$ whenever $n_{ij}(\varepsilon)>0$. Informational directness at ruler $\varepsilon$, by contrast, is the condition that the resolved conditional law lies within the direct family of Eq.~\eqref{eq:direct-family}, equivalently that the higher-order residual vanishes. There is therefore no implication
\begin{equation}
A_{ij}=1
\;\not\Rightarrow\;
I_{\rm higher}(\varepsilon)=0.
\label{eq:anatomical-vs-informational}
\end{equation}
The left-hand side of Eq.~\eqref{eq:anatomical-vs-informational} is a statement about adjacency. The right-hand side is a statement about factorization of the conditional response.

A monosynaptic anatomical connection therefore need not correspond to an additive computational interaction at finer resolution. Multiple contacts from the same axon can occupy different dendritic locations, participate in different local nonlinear regimes, and interact with different subsets of the surrounding input \cite{london2005,polsky2004}. When those variables are resolved, the response can contain higher-order structure. When they are collapsed through the somatic projection of Eq.~\eqref{eq:soma-weight}, that structure can be hidden inside one effective edge. Anatomical directness consequently does not imply informational additivity.

A minimal counterfactual makes the distinction explicit. Let $x_j\in\{0,1\}$ denote activity arriving through the anatomical connection $j\rightarrow i$, and let $n\in\{0,1\}$ denote a resolved local dendritic context generated by other inputs. Consider
\begin{equation}
\operatorname{logit}P(Y=1\mid x_j,n)
=b+\alpha x_j+\beta n+\gamma x_jn.
\label{eq:anatomical-context-model}
\end{equation}
The anatomy is identical for every value of $\gamma$: in all cases $A_{ij}=1$. What changes is the informational structure of the response. If $\gamma=0$, Eq.~\eqref{eq:anatomical-context-model} reduces to $b+\alpha x_j+\beta n$ and therefore belongs to the direct family of Eq.~\eqref{eq:direct-family}. The anatomical edge contributes the same log-odds increment irrespective of the local context.

If $\gamma\neq0$, the same anatomical edge behaves differently. Its context-dependent contribution is
\begin{equation}
\begin{aligned}
\Delta_j(n)
=\operatorname{logit}P(Y=1\mid x_j=1,n) \\
-\operatorname{logit}P(Y=1\mid x_j=0,n) \\
=\alpha+\gamma n.
\label{eq:anatomical-context-increment}
\end{aligned}
\end{equation}
For $n=0$, the increment in Eq.~\eqref{eq:anatomical-context-increment} is $\alpha$; for $n=1$, it is $\alpha+\gamma$. Thus the effect assigned to neuron $j$ changes even though the anatomical edge itself has not. When all four configurations of $(x_j,n)$ occur with nonzero probability and $\gamma\neq0$, the interaction term $x_jn$ cannot be represented as a sum $u_j(x_j)+u_n(n)$. The true conditional law therefore lies outside the direct family, and the conditional-KL representation of $I_{\rm higher}$ in Eq.~\eqref{eq:Ihigher_KL} gives a positive higher-order residual.

The counterfactual isolates the point cleanly. Setting $\gamma=0$ and $\gamma\neq0$ does not add or remove the anatomical connection $j\rightarrow i$. It changes only whether the influence carried by that connection is separable from the local dendritic context. The same anatomical graph can therefore support an informationally direct regime and an informationally synergistic regime.

The converse point is equally important. The existence of a contact field does not prove that every biological connection exhibits large higher-order information in every operating regime. If local integration is approximately separable over the distribution of states being sampled, a direct model may remain accurate even at finer resolution. The claim is instead one of non-identifiability: observing an accurate scalar dependency at neuronal scale cannot determine whether the microscopic realization beneath it was itself scalar and additive or whether richer contact-level interactions were removed by the projection.

This is why \emph{one edge} and \emph{one computational degree of freedom} become equivalent only after choosing the coarse ruler. At contact resolution, a single neuron-to-neuron edge can unfold into the contact field of Eq.~\eqref{eq:contact-field}, distributed across spatially distinct local domains. At neuronal resolution, that field is compressed back into the effective weight of Eq.~\eqref{eq:soma-weight}. The topology of the computational description therefore changes with the level at which nodes and edges are defined.

\subsubsection{Sparsity is scale-dependent}

The same projection that changes informational directness also changes the graph on which sparsity is defined. To make this explicit, let $\mathcal S_i(\varepsilon)$ denote the set of postsynaptic subunits of neuron $i$ resolved at ruler $\varepsilon$, and let
$A_{ija}^{(\varepsilon)}\in\{0,1\}$ indicates whether presynaptic neuron $j$ makes at least one resolved contact onto subunit $a\in\mathcal S_i(\varepsilon)$. The corresponding cell-level adjacency is obtained by collapsing all resolved subunit-level edges between the same neuronal pair:
\begin{equation}
A_{ij}^{\rm cell}
=
\mathbf{1}\!\left[
\sum_{a\in\mathcal S_i(\varepsilon)}
A_{ija}^{(\varepsilon)}>0
\right].
\label{eq:cell-adjacency-projection}
\end{equation}
Equation~\eqref{eq:cell-adjacency-projection} is itself a coarse-graining map. Several distinct presynaptic-to-subunit relations can produce the same binary statement $A_{ij}^{\rm cell}=1$.

The number of edges at the resolved and cell-level rulers is therefore
\begin{equation}
|E_\varepsilon|
=
\sum_{i,j}\sum_{a\in\mathcal S_i(\varepsilon)}
A_{ija}^{(\varepsilon)},
\qquad
|E_{\rm cell}|
=
\sum_{i,j}A_{ij}^{\rm cell}.
\label{eq:scale-dependent-edge-count}
\end{equation}
Because every cell-level edge requires at least one resolved subunit-level edge, Eq.~\eqref{eq:scale-dependent-edge-count} implies
\begin{equation}
|E_\varepsilon|\geq |E_{\rm cell}|.
\label{eq:edge-count-projection-bound}
\end{equation}
Equality in Eq.~\eqref{eq:edge-count-projection-bound} occurs only when every connected neuronal pair is represented by exactly one resolved postsynaptic subunit. Whenever a presynaptic neuron contacts more than one resolved subunit of the same postsynaptic cell, multiple fine-scale edges collapse onto a single cell-level edge.

The amount of topological compression can be quantified by

\begin{equation}
\chi(\varepsilon)
=
 \frac{|E_\varepsilon|}{|E_{\rm cell}|}
=
\frac{
\sum_{i,j}\sum_{a\in\mathcal S_i(\varepsilon)}
A_{ija}^{(\varepsilon)}
}{
\sum_{i,j}A_{ij}^{\rm cell}
}.
\label{eq:topological-compression-factor}
\end{equation}
The compression factor in Eq.~\eqref{eq:topological-compression-factor} is the mean number of resolved presynaptic-to-subunit edges represented by one connected neuronal pair. At the cell ruler, all of those relations are replaced by the single binary edge $A_{ij}^{\rm cell}=1$.

This makes the scale dependence of sparsity precise. Let $N_{\rm src}$ denote the number of possible presynaptic cells and $N_{\rm tgt}$ the number of postsynaptic cells. The cell-level edge density is
\begin{equation}
\rho_{\rm cell}
=
\frac{|E_{\rm cell}|}
{N_{\rm src}N_{\rm tgt}}.
\label{eq:cell-level-edge-density}
\end{equation}
At ruler $\varepsilon$, the resolved graph instead has
\begin{equation}
\rho_\varepsilon
=
\frac{|E_\varepsilon|}
{\displaystyle
N_{\rm src}\sum_i|\mathcal S_i(\varepsilon)|}.
\label{eq:resolved-edge-density}
\end{equation}

Equations~\eqref{eq:cell-level-edge-density} and \eqref{eq:resolved-edge-density} are densities on different graph spaces: the denominator itself changes when a postsynaptic cell is replaced by multiple resolved subunits. The two densities are nevertheless related exactly. Writing
$\bar s(\varepsilon)=N_{\rm tgt}^{-1}\sum_i|\mathcal S_i(\varepsilon)|$
for the mean number of resolved subunits per postsynaptic cell, Eqs.~\eqref{eq:topological-compression-factor}--\eqref{eq:resolved-edge-density} give
\begin{equation}
\frac{\rho_\varepsilon}{\rho_{\rm cell}}
=
\frac{\chi(\varepsilon)}{\bar s(\varepsilon)}.
\label{eq:density-ratio}
\end{equation}
Refinement therefore lowers the edge density whenever a connected presynaptic cell reaches fewer resolved subunits than the postsynaptic cell possesses. A bound follows because one presynaptic cell can contact at most $|\mathcal S_i(\varepsilon)|$ subunits of cell $i$: summing $\sum_a A_{ija}^{(\varepsilon)}\leq A_{ij}^{\rm cell}|\mathcal S_i(\varepsilon)|$ over all pairs gives
$\chi(\varepsilon)\leq \bar s_k(\varepsilon)$, where
$\bar s_k(\varepsilon)=\sum_i k_i^{\rm cell}|\mathcal S_i(\varepsilon)|/\sum_i k_i^{\rm cell}$
is the in-degree-weighted mean subunit count. Hence
\begin{equation}
\frac{\rho_\varepsilon}{\rho_{\rm cell}}
\leq
\frac{\bar s_k(\varepsilon)}{\bar s(\varepsilon)},
\label{eq:density-bound}
\end{equation}
with equality only when every connected pair contacts every resolved subunit. If the subunit count is the same for all cells, the right-hand side is unity and the resolved graph cannot be denser than the cell graph: the edge count cannot decrease by Eq.~\eqref{eq:edge-count-projection-bound}, while the density decreases whenever $\chi(\varepsilon)<\bar s(\varepsilon)$. Only a positive correlation between in-degree and subunit count can reverse this. Multi-contact connections that cluster on a few branches of an arbor with many branches correspond to $\chi(\varepsilon)\ll\bar s(\varepsilon)$, the regime in which the resolved graph is far sparser than its cell-level projection while carrying many more edges. Sparsity is not an invariant of the physical network; it is a property of the graph induced by the chosen ruler, and Eq.~\eqref{eq:density-ratio} states how it changes with that ruler.

The same distinction applies to degree. The cell-level in-degree of neuron $i$ is
\begin{equation}
k_i^{\rm cell}
=
\sum_j A_{ij}^{\rm cell},
\label{eq:cell-level-indegree}
\end{equation}
whereas the number of resolved presynaptic-to-subunit relations carried by that same cell is
\begin{equation}
k_i^{(\varepsilon)}
=
\sum_j\sum_{a\in\mathcal S_i(\varepsilon)}
A_{ija}^{(\varepsilon)}.
\label{eq:resolved-indegree}
\end{equation}
The two degrees count different objects. Equation~\eqref{eq:cell-level-indegree} counts distinct presynaptic cells; Eq.~\eqref{eq:resolved-indegree} counts distinct presynaptic-to-subunit relations. The difference between them is precisely the within-cell topology removed by the projection in Eq.~\eqref{eq:cell-adjacency-projection}.

A network can therefore be sparse as an axon-to-cell graph while possessing a substantially richer contact-to-subunit graph. Conversely, resolving additional contacts or subunits does not guarantee improved prediction: coarse descriptions can generalize better when data are finite and the fine degrees of freedom are poorly constrained \cite{herz2006}. What changes with resolution is the graph, the approximation space, and the set of interactions available to the model. Sparsity, degree, directness, and capacity must therefore be indexed by the ruler at which nodes and edges are defined.

The scalar synaptic weight is thus not a mathematical mistake. The mistake is to forget that it is an effective variable. The contact field in Eq.~\eqref{eq:contact-field} describes what may exist beneath the edge, while the somatic projection in Eq.~\eqref{eq:soma-weight} describes the operation required to make that fine structure look like one number. The coastline argument enters at exactly this point: a simple edge at one resolution does not imply a simple connection at every resolution.

\subsection{Resolution Expands An Approximation Space}

Having established that projecting a spatially distributed contact field onto a scalar weight can remove or re-express higher-order interaction structure, we now invert that perspective. Instead of asking what structure becomes inaccessible when fine variables are projected away, we ask what becomes representable when those variables are progressively resolved.

Here we map the complex physical geometry of a neuron onto the mathematics of approximation theory developed above. A model can only construct functions from the variables and nonlinear features made available at its chosen ruler. A point-neuron description that exposes only a few input coordinates therefore operates in a smaller approximation space than a description that additionally resolves nonlinear combinations generated within distinct subunits. Refining the ruler by un-pooling distinct dendritic subunits does not change the target function. It enlarges the feature space in which that same target can be approximated.

\begin{figure*}
\centering
\includegraphics[width=0.95\linewidth]{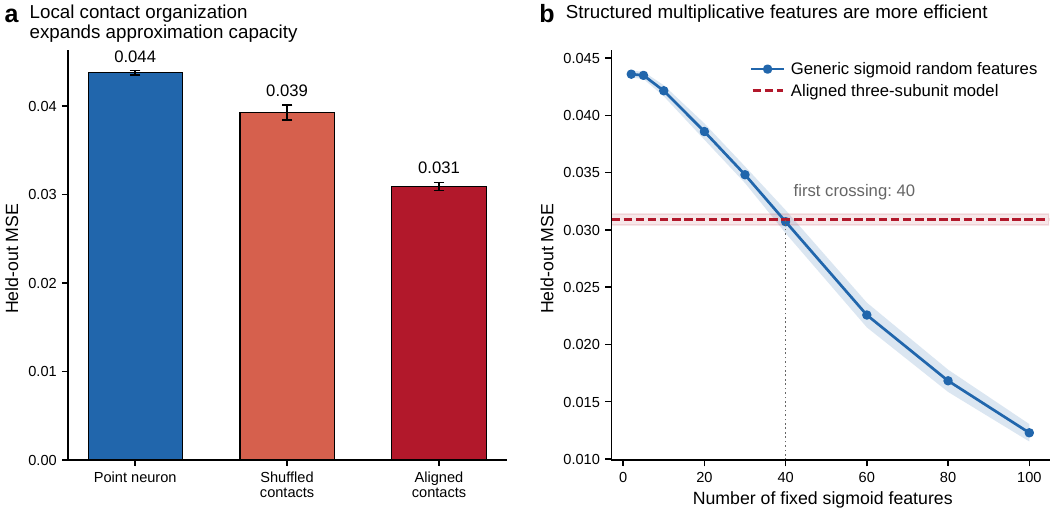}
\caption{\label{fig:capacity}\textbf{Organized multi-contact features improve held-out approximation.}
(a) Mean held-out MSE for a linear point neuron, a three-subunit model with shuffled contacts and a three-subunit model with contacts aligned to the target's pair structure.  Error bars are 95\% normal intervals across 100 independent train--test trials (6,000 samples in each split).  (b) Mean held-out MSE of a generic fixed sigmoid random-feature network.  The first tested width whose mean reaches the aligned model is \EquivalentHidden.  Shading shows 95\% normal intervals across 40 trials.}
\end{figure*}

This is the computational counterpart of the coastline argument. A coarse ruler does not prove that the underlying object has few degrees of freedom. Rather, it restricts the degrees of freedom available to the description. Conversely, resolving additional subunits can expose interaction terms that were absent from the coarse representation. The following constructions test that claim in two stages. We first compare point, shuffled-contact, and spatially organized models at fixed contact count. We then construct a genuinely nested hierarchy in which the same architecture is viewed at progressively finer resolutions.

This refinement also connects directly to the information-theoretic
formalism developed above.  The distinction between direct and
higher-order information is defined relative to the coordinates made
available to the model.  For example, the interaction $x_1x_2$ is a
joint dependence when the resolved coordinates are
$(x_1,x_2,\ldots)$, but becomes a one-coordinate dependence if the
refined description explicitly contains the subunit variable
$z_{12}=x_1x_2$.  Refinement can therefore move structure that appears
in $I_{\rm higher}$ at a coarse ruler into $I_{\rm dir}$ at a finer
one, without changing the underlying input--output law.

\subsubsection{The target function and methodological Safeguards}

We begin with six independent continuous inputs,
$\bm x\sim U[-1,1]^6$, and define the nonlinear target
\begin{equation}
f(\bm x)=\sigma\{2.5[x_{1}x_{2}+x_{3}x_{4}+x_{5}x_{6}+1.2x_{1}x_{3}x_{5}]-1.8\}
\label{eq:capacity-target}
\end{equation}
Equation~\eqref{eq:capacity-target} contains three pairwise products,
$x_1x_2$, $x_3x_4$, and $x_5x_6$, together with the third-order term
$x_1x_3x_5$. The target is therefore deliberately poorly matched to a model whose only features are independent first-order coordinates. It is not meant as a biophysical transfer function for a particular neuron. It provides a controlled function whose interaction structure is known before any model is fitted.

The centered input distribution ($U[-1,1]$ rather than $U[0,1]$) is an important methodological safeguard. For independent variables with zero mean, the raw product terms are orthogonal to the corresponding first-order coordinates under the sampling distribution. For example, the correlation between $x_1$ and $x_1x_2$ contains the factor $\mathbb E[x_2]=0$, and the same argument applies to the third-order term because the remaining factors also have zero mean. A linear model therefore cannot obtain an artificial advantage merely because all inputs and all products share the same positive offset.

This would not be true for inputs drawn only from a positive interval. If $x_i>0$ systematically, products such as $x_1x_2$ carry positive first-order correlations, allowing a linear readout to recover part of a nominally multiplicative target from marginal amplitude alone. Centering does not make Eq.~\eqref{eq:capacity-target} mathematically invisible to every possible low-order statistic, especially after the outer sigmoid. It does remove this elementary source of leakage and makes the comparison between first-order and interaction-sensitive representations substantially cleaner. By centering the inputs symmetrically around zero, the linear coordinates are orthogonal to the raw product terms, preventing the point model from exploiting mean-induced first-order correlations with those interactions. 

All models are evaluated on held-out samples rather than on their training error. The point of the experiment is therefore not whether a sufficiently flexible model can interpolate the target, but how efficiently different representations approximate it out of sample.

\subsubsection{Isolating geometric structure from contact count}

The first comparison asks whether exposing local nonlinear structure improves approximation when the number of underlying contacts is controlled. The coarse model is a ridge-fitted linear point readout whose features are only the six original input coordinates. Across 100 independent train--test trials, with 6,000 samples in each split, this model obtains mean held-out MSE \PointMSE.

We then refine the representation while preserving the same six axonal inputs. Each axon is represented by three contacts, giving 18 contacts in total, and those contacts are grouped into three nonlinear local subunits. In the aligned construction, the grouping respects the three pairwise terms in Eq.~\eqref{eq:capacity-target}: contacts associated with $x_1$ and $x_2$ occupy one subunit, those associated with $x_3$ and $x_4$ occupy another, and those associated with $x_5$ and $x_6$ occupy the third. Each subunit therefore supplies the readout with a local nonlinear feature capable of expressing coincidence structure that is absent from the six-dimensional linear point representation.

In the language of Eqs.~\eqref{eq:Idir}--\eqref{eq:Ihigher_KL},
the refinement does not create the interaction; rather, it changes whether
that interaction must be represented as a higher-order residual or is
available as a direct coordinate of the effective description.

The qualification ``pairwise'' matters. The aligned three-subunit architecture is designed around the three pair products in Eq.~\eqref{eq:capacity-target}. This is not handed the third-order term $x_1x_3x_5$ explicitly. The reduction in error therefore does not result from simply inserting every term of the target into the feature matrix. Rather, the experiment asks whether a small collection of appropriately organized local nonlinear features gives a better approximation basis than the unresolved point representation.

The results are positive. The aligned three-subunit model reduces the mean held-out MSE from \PointMSE\ to \AlignedMSE\ (Fig.~\ref{fig:capacity}a). The same six axons are present in both descriptions. What has changed is the resolution at which their local combinations are exposed to the readout.

Contact count alone, however, cannot establish that this improvement comes from organization. We therefore construct a shuffled control with the same number of axonal inputs, the same number of contacts, and the same number of nonlinear subunits, while randomizing which contacts occupy the individual subunits. The shuffled architecture retains the additional nonlinear machinery but destroys its systematic alignment with the pair structure of Eq.~\eqref{eq:capacity-target}. Its mean held-out MSE is \ShuffledMSE, intermediate between the point and aligned models.

The ordering
$\PointMSE>\ShuffledMSE>\AlignedMSE$
separates two effects. Merely exposing additional nonlinear features improves on the point representation, but organizing those features relative to the structure of the target improves it further. The approximation gain is therefore not explained solely by adding contacts or parameters. The arrangement of those contacts determines which interaction directions become readily available to the readout.

This distinction is closely related to the biological motivation of the paper. A neuron with many synapses does not obtain its computational repertoire from synapse count alone. Which inputs converge onto the same local nonlinear domain determines which coincidences can be represented economically. Contact number specifies how many microscopic elements are available; spatial organization specifies the basis functions those elements can collectively generate.

Figure~\ref{fig:capacity}b provides a second control on representational efficiency. We compare the organized three-subunit representation with a generic network of fixed sigmoid random features. As the random-feature width increases through the tested values, approximation improves, but the first tested width whose mean held-out error reaches the aligned three-subunit model is \EquivalentHidden\ features. This is not a lower bound on the number of units required by arbitrary neural networks, nor does it establish that dendritic features are universally optimal. It is a comparison within the specified feature families. Its significance is narrower: for this structured target, three appropriately organized local nonlinear features provide an economical basis compared with an unstructured random nonlinear expansion.

\subsubsection{Nested resolution and the participation ratio}

The preceding comparison changes the representation across three model classes. To isolate resolution more directly, we next construct a nested hierarchy in which the underlying nonlinear architecture is held fixed while progressively more of it is exposed.

For each trial, a 24-subunit sigma--pi architecture is drawn once. Each nonlinear subunit combines a subset of the available contacts through pairwise and third-order multiplicative terms followed by a local nonlinearity. Resolution is then varied without redrawing that architecture: the $n=1$ representation exposes the first subunit, the next resolution exposes the first two, and so forth through the tested sequence
$n=1,2,3,4,6,8,12,16,24$. Importantly, the original six direct input coordinates remain available at every resolution. Refinement therefore adds nonlinear features to an existing representation rather than replacing one set of inputs by another.

This nesting gives the word ``resolution'' a precise mathematical meaning. If the feature vector at resolution $n$ contains the six direct coordinates and the first $n$ nonlinear subunits, then every coarser feature set is contained in every finer one. Writing this nested representation as $\Phi_n(\bm x)\subset\Phi_{n+1}(\bm x)$ makes the hierarchy explicit: $\Phi_n$ is obtained from $\Phi_{n+1}$ by discarding resolved nonlinear coordinates.

Importantly, the original six inputs $\bm x$ remain present at every resolution, while every added subunit is itself a deterministic function of $\bm x$. Consequently,
$I(Y;\Phi_n)=I(Y;\bm x)$
for every $n$: refinement does not create additional information about the target. What increases is the approximation space available to the restricted readout. If $\mathcal H_n$ denotes the family of linear functions of $\Phi_n$, then $\mathcal H_n\subseteq\mathcal H_{n+1}$. Increasing $n$ therefore enlarges the hypothesis class without changing the underlying input--output information. Any improvement in held-out prediction can then be attributed to representational access rather than to an increase in $I_{\rm avail}$, and it can be interpreted against an explicitly nested hierarchy rather than against independently resampled architectures.

\begin{figure*}
\centering
\includegraphics[width=0.95\linewidth]{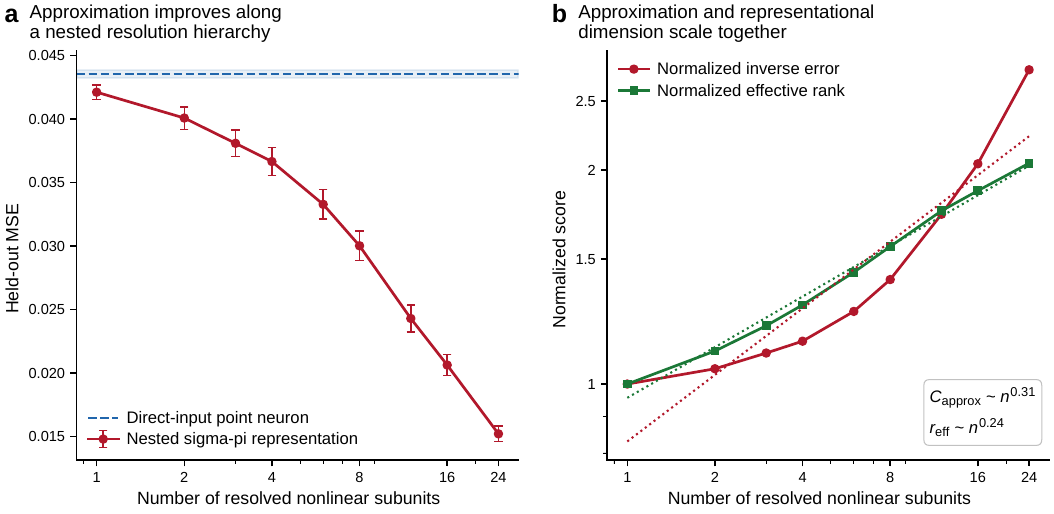}
\caption{\label{fig:scaling}\textbf{Approximation and effective dimension grow in a nested hierarchy.}
(a) Mean held-out MSE as progressively more nonlinear sigma--pi subunits are exposed from the same architecture; the dashed line is the direct-input point baseline.  Error bars and baseline shading are 95\% normal intervals across 60 independent architectures and train--test sets (5,000 samples per split).  (b) Normalized inverse error and normalized effective rank with log-space power-law fits.  Exponent intervals are 95\% percentile-bootstrap intervals over trials (2,000 resamples).}
\end{figure*}

Across 60 independently drawn architectures and train--test sets, the mean held-out MSE decreases monotonically over the tested resolutions, from \ScalingInitialMSE\ at one exposed nonlinear subunit to \ScalingFinalMSE\ at 24 (Fig.~\ref{fig:scaling}a). For comparison, the corresponding direct-input baseline, which contains the same six raw inputs but no nonlinear subunits, has mean MSE approximately $0.0436$.

To summarize the improvement, we define the normalized inverse-error score in the analysis as the error at the first resolved level divided by the error at resolution $n$, so that $C_{\rm approx}(1)=1$ and increasing values correspond to improved approximation. Over the tested range, its mean scaling is well described by
\begin{equation}
C_{\rm approx}\propto n^{\ApproxExponent},
\quad 95\%~{\rm CI}\ [\ApproxExponentLow,\ApproxExponentHigh],
\label{eq:approx-scaling}
\end{equation}
with a log--log fit exponent of approximately $0.31$. The relation in Eq.~\eqref{eq:approx-scaling} is an empirical scaling summary of this constructive hierarchy, not an exact law. In particular, its log-space $R^2$ is $0.8613$, so the purpose of the exponent is to quantify the systematic trend rather than to claim a universal power law for dendritic capacity.

Lower prediction error alone does not establish that the underlying representation has acquired independent dimensions; additional subunits could in principle supply largely redundant copies of features already present. We therefore measure the geometry of the representation itself, without reference to the target labels. Let $\lambda_k$ denote the nonzero eigenvalues of the covariance matrix of the centered feature representation. Its effective rank is defined by the participation ratio
\begin{equation}
r_{\rm eff}=\frac{(\sum_k\lambda_k)^2}{\sum_k\lambda_k^2}.
\label{eq:effective-rank}
\end{equation}
Equation~\eqref{eq:effective-rank} distinguishes nominal feature count from effective dimensionality. If variance is concentrated almost entirely along a small number of covariance directions, $r_{\rm eff}$ remains low even when many features are present. If variance is distributed across additional independent directions, the participation ratio rises.

The latter is what occurs in the nested hierarchy. Mean $r_{\rm eff}$ increases from approximately $6.86$ with one exposed subunit to $14.03$ with 24, and its normalized scaling is approximately $n^{\RankExponent}$, with 95\% interval
$[\RankExponentLow,\RankExponentHigh]$ (Fig.~\ref{fig:scaling}b). The corresponding log-space fit is particularly tight, with $R^2=0.9907$. The increase in nominal resolution is therefore accompanied by an increase in the effective dimensionality of the representation, not merely by accumulation of duplicate coordinates.

The two measurements answer complementary questions. $C_{\rm approx}$ in Eq.~\eqref{eq:approx-scaling} asks whether the enlarged representation provides a better basis for the particular target in Eq.~\eqref{eq:capacity-target}. The participation ratio in Eq.~\eqref{eq:effective-rank} asks whether that representation itself occupies a higher-dimensional space, independently of the target labels. Their joint increase is the relevant result. Explicitly, finer resolution both enlarges the accessible feature space and improves approximation within this controlled construction.

These exponents should not be interpreted as biological universality classes, nor as consequences of a putative dendritic fractal dimension. They characterize this particular nested sigma--pi ensemble. The more general conclusion does not depend on their numerical values. Once nonlinear subunits are unresolved, the coarse model is confined to the approximation space generated by the variables that survive the projection. Resolving additional subunits exposes additional interaction coordinates and thereby enlarges that space. A point-node description can therefore be an excellent effective model while remaining computationally lower-dimensional than the biological object it represents.

This completes the inversion of the argument developed in Sections V and VI. Coarse-graining showed how fine interaction structure can disappear from an effective description. Refinement shows the complementary operation: restoring spatially organized nonlinear degrees of freedom expands the set of functions available to that description. What appears as unexplained variation at the coarse ruler can become representable structure once the appropriate internal coordinates are resolved.

%========================================================================
% DISCUSSION
%========================================================================
\section{Discussion}
\subsection{Effective descriptions need a scale}
The adequacy of a low-order statistical model in explaining observed neural activity is a joint property of the underlying generative mechanism and the ruler through which it is observed. Pairwise maximum-entropy models can explain collective activity in small, sparsely active population of neurons~\cite{Schneidman06}, but this measure permits accurate fits even when the underlying interactions are not pairwise~\cite{roudi2009}. Common input can be a source of higher-order correlations~\cite{macke2011}, and larger neural populations can have explicit higher-order terms~\cite{ganmor2011}. The finding that direct, pairwise dependencies explain much of the output entropy of recorded neurons~\cite{lynn2026} therefore establishes an effective description at an observation scale rather identifying the underlying fine-scale computational mechanism. It does not provide a window to the computation implemented by intracellular voltages, channel states and contact locations absent from that description.

Our prior work illustrate why both the observable and the measure scale matter. In human cortex, pairwise models that exclude inhibitory neurons substantially overestimate excitatory synchrony~\cite{zanoci2019}. In multielectrode recordings from cats, monkeys and humans, neural avalanche statistics differ between spikes and field-potential, cautioning against interpreting apparent scaling as a unique criticality signature of the underlying generative mechanism~\cite{dehghani2012avalanches}. Spike avalanches show no clear power-law scaling, and the apparent scaling of local-field-potential negative peaks is shared by positive peaks unrelated to firing. Temporal coarse-graining of excitatory and inhibitory ensemble activity reveals coordination that persists across scales, while surrogates that randomize the precise timing of spikes disrupt that structure~\cite{dehghani2025symmetry}. These observations motivate examining which relations survive a change of ruler, rather than assuming that either universal simplification or universal loss of organization follow coarse-graining.

\subsection{Projection and mechanistic non-identifiability}
Our information-theoretic formalism separates two consequences of coarse-graining. It allows us to discern information loss from changes in apparent interaction order. Data processing inequality guarantees that under no projection the available predictive information could increase; however, it does not constrain the interaction order at which the surviving dependence may manifest. A strictly third-order dependency may disguise as pairwise interaction under projection, with partial or complete information preservation depending on the map (Fig.~\ref{fig:projection}). Along the branch-to-soma cascade, the direct fraction $F$ increases while available information decreases and the total-entropy fraction $R$ falls (Fig.~\ref{fig:cascade}). This \emph{denominator trap} concerns the fraction of surviving information under coarse-graining captured by the direct family, not the total-entropy fraction $R$. Residual $I_{\rm higher}$ contraction is mathematically guaranteed under the closure condition of Eq.~\eqref{eq:closure}; without it, no description of universal interaction order is meaningful (Fig.~\ref{fig:projection-resolution}). This traces the boundary of effective theories for neural systems: integrating out degrees of freedom renormalizes the orders of interactions as well as their strengths. Eliminating a mediating variable generates higher-order couplings among those that remain, while merging variables into a collective coordinate can instead absorb an interaction into the limiting definition of that coordinate. Apparent interaction order is therefore a property of the retained coordinates and descriptive measure, rather than an invariant of the generative mechanism. 

What a high $R$ does bound can be stated exactly. From Eqs.~\eqref{eq:S_ineq} and \eqref{eq:FracR}, any additional variable $\bm U$ satisfies
\begin{equation}
\begin{aligned}
I(Y;\bm U\mid\bm X_\varepsilon)
&\leq H(Y\mid\bm X_\varepsilon)\\
&\leq (1-R)\,S_{\rm tot}.
\end{aligned}
\label{eq:residual-bound}
\end{equation}
With $R>0.9$, as reported for mouse hippocampus and visual cortex~\cite{lynn2026}, an unobserved (dendritic) variable can not add more than a tenth of the output entropy to prediction of the same binarized output under the same distribution of population states. The the low-$R$ model cannot account for the generative dynamics behind those recordings (Fig.~\ref{fig:cascade}) . The lossless projection provides a counterexample: if $s_1$ is treated as the output, the conditional law $\operatorname{logit}P(s_1=1\mid s_2,s_3)=2Ks_2s_3$ is not captured by the direct family when the inputs are resolved separately, yet it is captured when the inputs are represented by their product (see Eq.~\eqref{eq:three-spin} and Fig.~\ref{fig:projection}b). Increasing $K$ drives the fraction of its total-entropy towards unity without changing that distinction. Because analyzed extracellular population activity are themselves downstream of the outputs of other neurons in the network — each already a nonlinear function of its own afferents — they can correlate with the presynaptic pattern of the cell under study without being that pattern. A high $R$  among recorded variables limits the additional predictive information an unobserved variable can contribute to (Eq.~\eqref{eq:residual-bound}); however, it does not discern whether the neuron implements a linear summation of synaptic inputs or computes a nonlinear combination.

A recent work shows that fitting a maximum‑entropy model while constraining only a selected set of summary statistics is equivalent to projecting the empirical sampling distribution onto that restricted family \cite{safaai2026leakage}. Therefore, any higher‑order or nonlinear interactions that covaries with the constrained statistics cannot be represented explicitly; instead, it will be absorbed into the low‑order parameters. This misattribution— what they call ``feature leakage''—would be particularly pronounced in sparse, unevenly sampled binary data. When input sampling is sparse and biased, rare input combinations and biased sampling would cause different nonlinear/higher-order mechanisms to project onto the low‑order statistics, so they look like mere shifts in mean firing or pairwise covariance. To isolate this effect, they use a procedure that keeps each neuron’s conditional response distribution (the probability of each output given a specific input pattern) unchanged and  only alter the input‑state weights used in the fit, so any change in attribution must come from the sampling geometry. An example would be replacing empirical sampling weights with product‑marginal or balanced weights. Because the conditional responses are unchanged, it would follow that that any change in the fitted attribution must stem from the altered sampling geometry. 
In their examination of hipocampal recordings, the median fraction of explanatory power attributed to first‑order terms falls from 0.75 (empirical) to 0.68 (product‑marginal) and 0.46 (balanced). One can then conclude that the apparent low‑order explanatory power depends on which input states were over‑ or under‑sampled rather than the system harboring genuinely low‑order generative structure \cite{safaai2026leakage}.

Our construction unifies two apparently conflicting observations — that a low‑order fit can be predictively adequate under the observed state distribution \cite{lynn2026} and that the same fit can fail to identify the underlying conditional mechanism \cite{safaai2026leakage} — by showing they probe different arguments of the same object. 
The ``direct fraction'' is a function of three variables: i) the distribution over input states, ii) the measurement map that defines the observed coordinates, and iii) the model family used for fitting. One analysis computes the direct fraction at the empirical value of all three (distribution, observed coordinates, and chosen model), concluding predictive adequacy under those conditions \cite{lynn2026}. The other keeps the coordinates and model fixed but perturbs the sampling distribution, finding the attribution unstable and thereby revealing feature leakage \cite{safaai2026leakage}. We instead vary the measurement map and the fitted family at fixed mechanism and show that \emph{interaction order itself need not be preserved}. Leakage is the sampling‑geometry counterpart of the state‑space many‑to‑one measurement map. A concentrated input distribution folds higher‑order interactions into effective low‑order parameters in parameter space, just as a many‑to‑one measurement map converts interaction order in state space (Figs.~\ref{fig:projection}, \ref{fig:cascade}). The scale‑indexed decomposition provides the coordinate system in which both perspectives are visible simultaneously: changing the distribution, the observed coordinates, or the fitted family lawfully changes the direct fraction. Predictive adequacy and mechanistic non-identifiability are therefore two sides of one moon; the scale-indexed decomposition brings the dark side of that moon into view. 

\subsection{Biophysics links resolution to computation}
Membrane biophysics supplies the physical content that makes Eq.~\eqref{eq:sufficiency} a testable claim. For a deterministic projection $\bm Z=T(\bm X)$, lossless observational sufficiency for a fixed output $Y$ requires
\begin{equation}
\begin{aligned}
I(Y;\bm X\mid\bm Z)&=0,\\
\text{equivalently}\qquad
P(Y\mid\bm X)&=P(Y\mid\bm Z)
\end{aligned}
\label{eq:sufficiency}
\end{equation}
almost surely. 

The point-neuron hypothesis is the specific claim that this equality holds when $Z$ is a weighted sum of presynaptic activities, $Z=\sum_kw_kx_k$. It then follows that direct family in Eq.~\eqref{eq:direct-family} is a sufficient descriptor. In other words, the somatic drive can adequately predict the measured spiking output. Cable theory provides the bound when the hypothesis is justified. If the dendritic tree were passive, linearity and superposition would translate to each synaptic contribution being routed through a fixed transfer impedance. In that case, the somatic drive would be exactly equivalent to a weighted-sum of presynaptic currents, and the sufficiency condition would hold. Active dendritic conductances or nonlinear synaptic interactions violate superposition. This departure from passivity can invalidate the weighted-sum sufficiency even if a low-order fit appears to provide adequate statistical predictivity.  

Neuronal biophysics determines how spatially distributed inputs invoke local dendritic responses and turn into somatic signals. The projection $T_\varepsilon$ therefore finds a physical meaning: integration and propagation can make input configurations indistinguishable at one observational scale while remaining separable at another. Which distinctions are preserved or coarse-grained depends on membrane dynamics, electrotonic transfer, and the observer’s choice of recorded variables. Our construction isolates these transformations without assuming a particular neural morphology or network architecture.

The dependence on local context begins at the synaptic contact. Unlike prescribed current sources on a passive tree, conductance-based inputs do not generally superpose. The current through contact $k$ is
\begin{equation}
I_k=\bar g_k\,x_k\,M_k(V_{\rm loc})\,
\bigl(V_{{\rm rev},k}-V_{\rm loc}\bigr),
\label{eq:synaptic-current}
\end{equation}
where $V_{\rm loc}$ is local dendritic voltage and $V_{{\rm rev},k}$ the reversal potential. For voltage-independent synapses $M_k=1$. For NMDA receptors, $M$ describes relief of Mg$^{2+}$ block, reflecting that NMDA conductance is contingent on coordinated local depolarization~\cite{jahr1990}. Inputs from other synapses within an electrotonic neighborhood, that change $V_{\rm loc}$ of a given synapse, alter that synapse’s contribution. This is the physical basis for the context dependence formalized in Eq.~\eqref{eq:contact-context} and the term $\gamma x_jn$ of Eq.~\eqref{eq:anatomical-context-model}. Depolarization reduces the excitatory driving force, leading to progressively smaller contribution from additional inputs. Interestingly, the same depolarization can relieve NMDA block, thereby recruiting an amplified response to coordinated excitation \cite{polsky2004,palmer2014}.
Compression and coincidence detection therefore arise from distinct voltage dependencies of synaptic current and superposition of inputs. The point‑neuron hypothesis is therefore the weak‑input limit of the membrane equation: It is a regime, not a computational property, of the cell.

These mechanisms motivate the operations in the cascade (Fig.~\ref{fig:cascade}). The triple products in Eq.~\eqref{eq:cascade} schematize local coincidence detection. The cap in Eq.~\eqref{eq:branch-observables} illustrates how saturation can erase distinctions among input configurations, while weighted somatic mixing abstracts location-dependent electrotonic transfer~\cite{london2005}. As summary statistics, the branch and somatic variables reduce the description to a saturating count per branch and a weighted sum across branches. They do not represent literal membrane potentials through which the output is routed, and the output law does not factor through these variables. The construction tests the sufficiency of a point‑like reduction on a neuron whose output is driven by local coincidences. Nothing in membrane physics guarantees the validity of one scalar per branch, much less one per cell. Biophysical properties such as local voltage, NMDA‑receptor occupancy, and calcium dynamics are distinct state variables with different input dependencies and time constants. The point is to reveal what is lost when an observer makes fewer distinctions than the fine response law mandates.

The spatial ruler also has a biophysical root. Contacts share the local voltage ($V_{\rm loc}$) over a neighbourhood set by the electrotonic length constant ($\ell=\sqrt{dR_{\rm m}/4R_{\rm a}}$, where $d$ is dendritic diameter, and $R_{\rm m}$ and $R_{\rm a}$: the membrane and axial resistivities), and by the impedance mismatch at branch points. Together with branch-point impedance and active conductances, electrotonic coupling defines the functional domains represented by $\mathcal S_i(\varepsilon)$. 
Thin branches, having high input impedance and weak coupling to their neighbours, behave as nonlinear compartments~\cite{polsky2004,losonczy2006}. However, their effective independence depends on state: several-fold fall in effective membrane resistance can shorten electrotonic reach in high-conductance state of active cortex~\cite{destexhe2003}, and inhibition along the path to the soma can selectively suppress a branch contribution~\cite{koch1983}. These domains support distal--proximal coupling~\cite{larkum1999}, temporal sequence discrimination~\cite{branco2010} and branch-specific calcium spikes gate plasticity~\cite{cichon2015}. % dendritic events shape tuning and sensory-evoked spiking in vivo~\cite{lavzin2012,smith2013,palmer2014},  and even passive, sublinearly summing branches allow linearly non-separable computation~\cite{caze2013}.
The appropriate spatial ruler (the number of effectively independent subunits and the weights of the somatic mixture) therefore depends on the network state as well as the geometry of the excitable units.

Read in the other direction, the same physics explains why neuronal geometry and spatial organization of contacts is computationally load‑bearing. Scattering a multi‑contact connection across electrically semi‑independent compartments need not be anatomical redundancy; it can enlarge the set of nonlinear features available to the cell. 
Synaptic convergence onto a nonlinear compartment invokes local coincident activity \cite{polsky2004,gidon2020,beniaguev2021}. In contrast, scattered input across compartments preserves distinct response and limits shared saturation \cite{london2005}. Therefore, how the inputs share compartments determine what nonlinear combination of the inputs can be economically computed by the neuron \cite{poirazi2003}. Multicontact segregation therefore enables the retention of higher‑order cross‑terms as distinct features that a single summation node would otherwise erase. This is the biophysical counterpart of the approximation-space argument: organized subunits expose useful features (Fig.~\ref{fig:capacity}), and resolving additional subunits can enlarge the computational space accessible to a restricted readout (Fig.~\ref{fig:scaling}). Those transformations are determined by morphology, conductances and contact placement rather than by contact count alone. Compressing each neuron's computation to a node and each connection to a scalar edge removes contact-to-subunit assignments unless they are compensated through additional state or effective parameters. Cell-level connectivity therefore constrains the circuit but does not fully determine the computation performed at its nodes.

\subsection{Computational scale in neuro-inspired network design}
Our approximation results show why simply matching neuron counts and weight estimates is insufficient for  principled neuro‑inspired design. In our comparison, three aligned subunits can reach the accuracy that forty fixed sigmoid random features (Fig.~\ref{fig:capacity}). Shuffling the contacts degrades the network performance despite preserving contact count and subunit count. The gain therefore depends on which inputs share a nonlinear operation. The nested hierarchy makes this effect explicit; $I(Y;\Phi_n)=I(Y;\bm x)$ remains fixed, while $\mathcal H_n\subseteq\mathcal H_{n+1}$ expands the readout's approximation space. The result is prediction improvement and an enhanced effective rank as resolution exposes useful higher-order interactions (Fig.~\ref{fig:scaling}). These results identify a concrete architectural choice. One options is to harness useful interactions by by adding variables inside each unit so those computations remain local. Alternatively, one can keep units simple and rely on a larger network to reconstruct the same interactions. A biological neuron and an artificial unit need not occupy the same computational scale.

The same compression issue arises when one tries to abstract the connectome wiring and map it into a network design. Equation~\eqref{eq:cell-adjacency-projection} collapses distinct presynaptic-to-subunit relations into one cell-level edge. The compression factor $\chi(\varepsilon)$ counts that multiplicity, and Eq.~\eqref{eq:density-ratio} shows that apparent sparsity depends on the chosen graph. A biological connection composed of multiple synaptic contacts therefore carries organization detail that a cell-level adjacency does not specify. Adjacency alone does not determine functional influence. Equation~\eqref{eq:soma-weight} makes the effective weight a function of the contact field and local state. And in Eq.~\eqref{eq:anatomical-context-increment}, the same anatomical edge contributes $\alpha+\gamma n$ as dendritic context changes. Morphology, contact placement and membrane state thus sit beneath the graph and shape computation. Connectome-constrained models reproduce fly visual tuning \cite{lappalainen2024} and sensorimotor responses \cite{shiu2024}. Their success supports the tested cell-level descriptions, but it does not recover the within-cell organization lost via projection.

Functional connectomics can provide a practical route from biological measurements to network design if the extracted priors reflect scale-dependent measurements. Using MICrONS (Machine Intelligence from Cortical Networks) data~\cite{bae2025functionalconnectomicsspanning}, we have shown how activity-derived relationships and cortical geometry can guide recurrent networks that outperform randomly initialized baselines \cite{shakiba2026harnessing}. We used functional relationships to initialize recurrent weights, neuronal positions to define spatial costs, and anatomical connectivity to constrain network communicability. These constraints improve task performance. Functional initialization provides the largest performance gain, while measured geometry adds a further benefit without increasing model size~\cite{shakiba2026harnessing}. Our scale-indexed decomposition framework extends this design question to dendritic structure, where contact-to-subunit assignments determine which nonlinear features become accessible.

The biophysical models provide concrete implementations of the interactions our theory highlights. The voltage-dependent NMDA factor in Eq.~\eqref{eq:synaptic-current} couples each contact to local depolarization, enabling context-dependent integration. In detailed model of pyramidal neurons, optimizing synaptic strengths and locations can enable reasonable XOR and ten-bit parity performance~\cite{aizenbud2026compute}. Removing NMDA conductances or dendritic structure impairs that capability. These are in-silico model demonstrations, not evidence of in-vivo computation. However, they illustrate that computational power depends on internal organization suppressed by single‑node descriptions. A bilinear reduction captures part of that organization through location- and timing-dependent pairwise coefficients \cite{li2019}. Seen through the lens of our framework, such approach enlarges the effective family within its tested subthreshold regime. 

For neuro-inspired design, the question is therefore explicit: \emph{which contact groups, nonlinear features and state dependencies should each model unit retain?} The same question applies when reducing circuit motifs for composition. We have shown that coupled models of canonical circuit motifs (divisive competition and winner-take-all normalization) can generate joint computational operations absent from the uncoupled motifs~\cite{dehghani2026moredifferent}. A reduction adequate for an isolated motif must therefore be reassessed when that motif interacts with others, answering \emph{does the composition preserve the state dependencies through which the coupling acts?} Our scale-indexed decomposition and computational results provide controlled comparisons that answer the question of what to retain, beyond merely matching cell counts or copying cell-level connectivity.

\subsection{How long is, then, the coast of a neuron?}
In analogy with Mandelbrot’s coastline paradox ~\cite{mandelbrot1967}, the question is well posed only after specifying i) the ruler, ii) the target observable, iii) the class of perturbations, and iv) the domain of conditions. Our scale‑dependent construction makes mechanistic non‑identifiability explicit and shows why the amount of information lost at a given scale in a biological regime must be determined empirically. One should hold the output fixed and then ask how much extra predictive information finer‑grained variables contribute. That comparison is a direct test of observational sufficiency (Eq.~\eqref{eq:sufficiency}). Doing so isolates the contribution of higher‑order interactions and finer variables from the residual that any fitted model family leaves unexplained. One can then apply perturbations (such as branch‑targeted inhibition, shifts in spatial input clustering, or changes to the temporal input regime) to probe whether an observational description that was adequate continues to be so under new conditions. The continued success across such tests supports the validity of reduction and corase-graining within the tested domain. Such empirical support, however, does not uniquely identify the microscopic implementation.

A coarse variable can be simple to record, easy to fit, and straightforward to simulate. It may even account for most of the in‑distribution output entropy. Yet that convenience can be misleading. The same somatic or point description that suffices to forecast spikes in a fixed population regime may fail to capture phenomena that depend on subcellular detail. Branch voltages, for example, may be essential to explain coincidence detection or distal–proximal coupling. Likewise, contact‑ or spine‑resolved variables are often required for questions about plasticity.

Conversely, resolving more microscopic detail can increase estimation error without improving the prediction of interest~\cite{herz2006}. Interaction order, directness, sparsity, and capacity are joint properties of the neuron and the chosen scale of description. Integrating out hidden degrees of freedom renormalizes apparent interaction strengths and can change which orders of interaction appear dominant. The appropriate scale is therefore question-dependent. Neither tractability nor goodness of fit alone establishes that a coarse-graining preserves the quantities an inference about computation requires. The point-neuron is then a ruler. It is a useful, predictive abstraction. However, it becomes misleading when its readings are reported as properties of the coast.

%======================

% ----------------------------------------------------------------------
% Acknowledgement
% ----------------------------------------------------------------------
\begin{acknowledgments}
I wish to thank Gregory Chaitin for extended and deeply illuminating discussions on scale and information complexity, which substantially shaped the conceptual development of this study. I am also grateful to colleagues at the McGovern Institute and IAIFI for conversations on scale‑dependent computation and dendritic morphology.
\end{acknowledgments}

% ----------------------------------------------------------------------
% Data/Code
% ----------------------------------------------------------------------

\section*{Code Availability}
Code implementing the model, exact and held-out numerical results, source data for plotted series and figure are available at: \\ \url{https://github.com/neurovium/CoastOfNeuron}.

%======================
% Refs
%======================

\section*{References}
\bibliography{refs}

%======================
% APPENDIX
%======================
\newpage
\appendix

\section{ALGEBRAIC DERIVATIONS AND EXACT CALCULATIONS}

\subsection{Expansion of the interaction residual}
The higher-order interaction residual is defined as the difference between the direct conditional cross-entropy and the true conditional entropy: $I_{higher} = S_{dir} - S_{true}$. Expanding this into its constituent expectations yields the conditional Kullback-Leibler divergence. Let $P \equiv P_{\epsilon}(Y|X_{\epsilon})$ and $Q^{*} \equiv Q_{\epsilon}^{*}(Y|X_{\epsilon})$:

\begin{equation}
\begin{aligned}
I_{higher} &= \left( -\mathbb{E}_{P_{\epsilon}} \log_{2} Q^{*} \right)
              - \left( -\mathbb{E}_{P_{\epsilon}} \log_{2} P \right) \\
&= \mathbb{E}_{X_{\epsilon}} \Bigg[
      \sum_{y} P(y|x) \log_{2} P(y|x) \\
&\qquad\qquad
      - \sum_{y} P(y|x) \log_{2} Q^{*}(y|x)
    \Bigg] \\
&= \mathbb{E}_{X_{\epsilon}} \left[
      \sum_{y} P(y|x)
      \log_{2} \left( \frac{P(y|x)}{Q^{*}(y|x)} \right)
    \right] \\
&= \mathbb{E}_{X_{\epsilon}}
    D_{KL}[P_{\epsilon}(Y|X_{\epsilon}) || Q_{\epsilon}^{*}(Y|X_{\epsilon})]
\end{aligned}
\end{equation}

% \begin{equation}
% \begin{aligned}
% I_{higher} &= \left( -\mathbb{E}_{P_{\epsilon}} \log_{2} Q^{*} \right) - \left( -\mathbb{E}_{P_{\epsilon}} \log_{2} P \right) \\
% &= \mathbb{E}_{X_{\epsilon}} \left[ \sum_{y} P(y|x) \log_{2} P(y|x) - \sum_{y} P(y|x) \log_{2} Q^{*}(y|x) \right] \\
% &= \mathbb{E}_{X_{\epsilon}} \left[ \sum_{y} P(y|x) \log_{2} \left( \frac{P(y|x)}{Q^{*}(y|x)} \right) \right] \\
% &= \mathbb{E}_{X_{\epsilon}} D_{KL}[P_{\epsilon}(Y|X_{\epsilon}) || Q_{\epsilon}^{*}(Y|X_{\epsilon})]
% \end{aligned}
% \end{equation}
% By Gibbs' inequality, $D_{KL} \ge 0$, mathematically guaranteeing $I_{avail} \ge I_{dir}$.

\subsection{Joint surrogate and KL contraction}
To bound the higher-order information under a coarse-graining map $Z=T(X)$, we construct the joint surrogate $Q_{Y,X}^{*}(y,x) = P(x)Q^{*}(y|x)$. The data-processing inequality (DPI) for KL divergence states that applying the same deterministic map $T$ to two distributions guarantees contraction. If the projected surrogate $Q_{Y,Z}^{*}$ remains within the valid direct family at the coarse scale, we obtain the chain:
% \begin{equation}
% I_{higher}(Z) \le D_{KL}(P_{Y,Z} || Q_{Y,Z}^{*}) \le D_{KL}(P_{Y,X} || Q_{Y,X}^{*}) = I_{higher}(X)
% \end{equation}

\begin{equation}
\begin{aligned}
I_{\text{higher}}(Z)
&\le D_{KL}(P_{Y,Z} \,\|\, Q_{Y,Z}^{*}) \\
&\le D_{KL}(P_{Y,X} \,\|\, Q_{Y,X}^{*})
= I_{\text{higher}}(X)
\end{aligned}
\end{equation}

The first inequality accounts for the potential re-optimization of the direct family at the coarse scale, while the second inequality is the strict DPI contraction, ensuring the absolute synergistic residual cannot grow.

\subsection{Partition function and MaxEnt marginalization}
For the three-spin system, the true joint distribution is driven strictly by the triad coupling: $P_{K}(s) \propto \exp(K s_{1}s_{2}s_{3})$. The normalization constant (partition function) $Z$ requires summing over all $2^{3}=8$ states. Four states yield $s_{1}s_{2}s_{3}=+1$ and four yield $-1$:
\begin{equation}
Z = 4\exp(K) + 4\exp(-K) = 8 \cosh K
\end{equation}
To determine the pairwise maximum-entropy projection $\tilde{P}^{(2)}$, we evaluate the lower-order moments. Marginalizing over $s_{3}$ to find the pairwise probability $P(s_{1}, s_{2})$ yields:
\begin{equation}
\begin{aligned}
P(s_{1}, s_{2}) &= \sum_{s_{3} \in \{-1, 1\}} \frac{\exp(K s_{1}s_{2}s_{3})}{8 \cosh K} \\
&= \frac{\exp(K s_{1}s_{2}) + \exp(-K s_{1}s_{2})}{8 \cosh K} = \frac{2 \cosh(K s_{1}s_{2})}{8 \cosh K}
\end{aligned}
\end{equation}
Because $\cosh$ is an even function, $\cosh(K) = \cosh(-K)$, and the numerator is $2 \cosh K$ for all four possible $(s_{1}, s_{2})$ pairs. Thus, $P(s_{1}, s_{2}) = 1/4$. All 1st and 2nd moments are strictly zero. The unique maximum-entropy distribution $\tilde{P}^{(2)}$ matching zero constraints is the independent uniform distribution, yielding an entropy of exactly $H[\tilde{P}^{(2)}] = 3 \log_{2}(2) = 3$ bits.

\subsection{Exact enumeration of the branch-to-soma cascade}
To explicitly track the loss of higher-order information across measurement scales, the cascade is evaluated via exact state enumeration over a uniform input distribution. There are $2^{6} = 64$ possible configurations of the 6 independent binary inputs, such that $P(x) = 1/64$ for all $x \in \{0,1\}^{6}$. This finite-state construction ensures that the expectations $\mathbb{E}_{P}$ for cross-entropy and mutual information are exact discrete sums, avoiding Monte Carlo sampling variance.

The data-processing inequality enforces a contraction of total available information as the cardinality of the observable state space ($|\mathcal{S}|$) shrinks under successive projections:
\begin{enumerate}
    \item \textbf{Level 1 (Synaptic):} Observables $X \in \{0,1\}^{6}$. $|\mathcal{S}_{1}| = 64$.
    \item \textbf{Level 2 (Branch):} Projection $B = (B_{1}, B_{2})$ where $B_{1} = \min(\sum_{1}^{3} x_{i}, 2)$ and $B_{2} = \min(\sum_{4}^{6} x_{i}, 2)$. State space $B \in \{0,1,2\}^{2}$. $|\mathcal{S}_{2}| = 9$.
    \item \textbf{Level 3 (Symmetric Soma):} Projection $Z_{sym} = 0.5 B_{1} + 0.5 B_{2}$. The observable states are $\{0, 0.5, 1.0, 1.5, 2.0\}$. $|\mathcal{S}_{3, sym}| = 5$.
    \item \textbf{Level 3 (Asymmetric Soma):} Projection $Z_{asym} = 0.6 B_{1} + 0.4 B_{2}$. The observable states are $\{0, 0.4, 0.6, 0.8, 1.0, 1.2, 1.4, 1.6, 2.0\}$. $|\mathcal{S}_{3, asym}| = 9$.
\end{enumerate}

To calculate the direct information $I_{dir}$, we optimize the direct surrogate $Q^{*}$ over two distinct hypothesis families. The \textit{linear-moment} family enforces a strict scalar weight on the observable: 
\begin{equation}
\log \left( \frac{Q^{*}(Y=1|Z)}{Q^{*}(Y=0|Z)} \right) = w_{0} + w_{1} Z
\end{equation}
Conversely, the \textit{categorical-additive} family avoids enforcing arbitrary linear distances between states by assigning independent weights via one-hot indicators. For the intermediate branch level, this is defined as:
% \begin{equation}
%     \log \left( \frac{Q^{*}(Y=1|B)}{Q^{*}(Y=0|B)} \right) = w_{0} + \sum_{k=1}^{2} w_{1,k} \mathbb{1}[B_{1}=k] + \sum_{k=1}^{2} w_{2,k} \mathbb{1}[B_{2}=k]
% \end{equation}

\begin{equation}
\begin{split}
\log\!\left( \frac{Q^{*}(Y=1|B)}{Q^{*}(Y=0|B)} \right)
= w_{0}
+ \sum_{k=1}^{2} w_{1,k}\,\mathbb{1}[B_{1}=k] \\
\quad
+ \sum_{k=1}^{2} w_{2,k}\,\mathbb{1}[B_{2}=k]
\end{split}
\end{equation}

Under the symmetric somatic map the categorical-additive residual is exactly zero for a structural reason: only one observed coordinate remains, and an unrestricted function of a single coordinate reproduces $P(Y\mid Z)$ exactly. The value $I_{\rm higher}=0$ therefore reflects the family and the number of retained coordinates, not additivity of the mechanism in Eq.~\eqref{eq:cascade}. What the symmetric map removes is branch identity: it merges the nine branch states into five, and $I_{\rm avail}$ falls from $0.028607$ to $0.023526$ bits. The asymmetric map with $\alpha=0.6$ is instead one-to-one on the nine branch states, so it retains all of the branch-level information, $I_{\rm avail}=0.028607$ bits. Its effect appears in the linear-moment family, which must describe nine states with a single slope in $Z$: $F$ falls from $98.9\%$ to $78.4\%$. The categorical family is fully direct for either scalar map by construction.

\section{NUMERICAL DETAILS}

\subsection{Exact information projections}

All discrete examples are evaluated by exact state enumeration; no Monte Carlo entropy estimator is used.  For an observable $\bm z$, states with identical $\bm z$ are aggregated to obtain $P(\bm z)$ and $P(Y=1\mid\bm z)$.  The direct model is fitted by minimizing the exactly weighted Bernoulli cross-entropy.  Binary coordinates enter linearly.  Multilevel coordinates use one-hot indicators for all but one reference level, implementing the univariate functions $u_i$ in Eq.~(\ref{eq:direct-family}).  Linear-moment results instead use the numerical observable itself.  Optimization uses L-BFGS-B with an analytic gradient.  The exported validator checks
\begin{equation}
 I_{\rm avail}=I_{\rm dir}+I_{\rm higher}
\end{equation}
to numerical tolerance and checks that every reported fraction lies in $[0,1]$.

\subsection{Projection resolution}

To expose dependence on the final ruler, the supplementary calculation uses
\begin{equation}
 Z_{\alpha,\Delta}
 =\Delta\,{\rm round}\!\left[
 \frac{\alpha B_1+(1-\alpha)B_2}{\Delta}\right],
\end{equation}
with $\alpha\in[0,1]$ and $\Delta\in[0.05,1]$.  The heat maps in Fig.~\ref{fig:projection-resolution} use the linear-moment direct family.  Some apparently fine quantizations retain all distinct weighted sums; others merge them.  The resulting bands show why no projection-independent monotonicity claim is available.

\renewcommand{\thefigure}{S\arabic{figure}}
\renewcommand{\theHfigure}{S\arabic{figure}}
\setcounter{figure}{0}
\begin{figure*}
\centering
\includegraphics[width=0.95\linewidth]{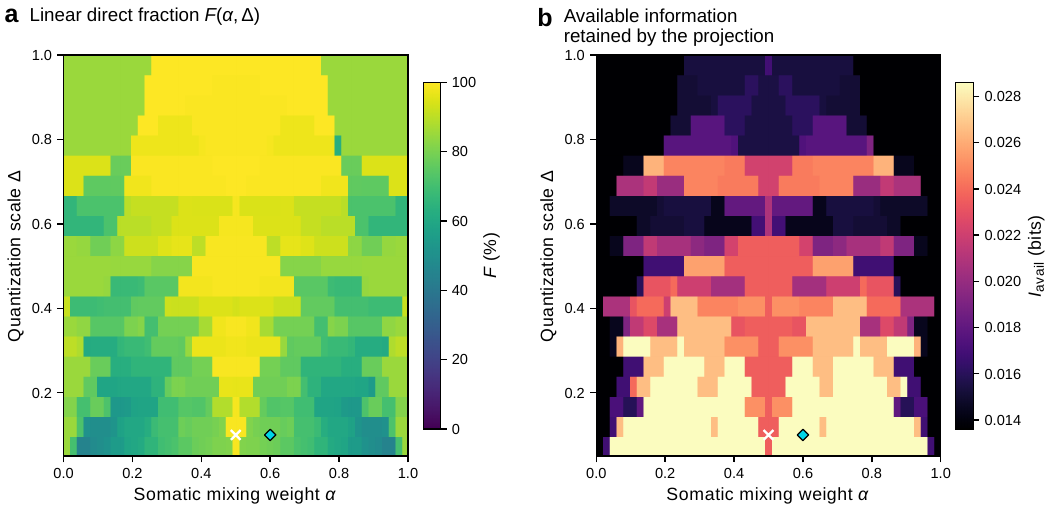}
\caption{\label{fig:projection-resolution}\textbf{The final direct fraction depends on mixing and measurement resolution.}
(a) Linear-moment $F$ for the exact cascade after weighted somatic mixing and quantization.  (b) Available information retained by the same projection.  At $\Delta=0.1$, the white cross marks symmetric $0.5/0.5$ mixing and the cyan diamond marks asymmetric $0.6/0.4$ mixing.  Heat-map values are exact; discontinuous bands arise when quantization merges different branch-state pairs.}
\end{figure*}

\subsection{Held-out capacity calculations}

All continuous inputs are independent and uniform on $[-1,1]$.  Linear readouts are ridge fits with an unpenalized intercept and predictions clipped to $[0,1]$.  Figure~\ref{fig:capacity} uses 100 independent trials for the three contact models and 40 independent trials for each random-feature width.  The random-feature model uses fixed sigmoid features within a trial and trains only its ridge readout.

For Fig.~\ref{fig:scaling}, each nonlinear subunit receives four contacts and contains weighted pair products plus one triple product followed by a hyperbolic tangent.  The feature set at $n$ subunits is a strict subset of the feature set at every larger sampled $n$.  Effective rank is computed from the eigenvalues of the held-out feature covariance.  Scaling exponents are ordinary least-squares fits in log space to normalized mean curves.  Confidence intervals for means are mean $\pm1.96$ standard errors; exponent intervals are percentile bootstraps over the 60 trials.  No morphology, cable equation or active-conductance parameter is fitted or implied.

\end{document}